\documentclass[12pt,aip,preprint,superscriptaddress,onecolumn,nolengthcheck]{revtex4-1}
\usepackage{graphicx}  
\usepackage{siunitx}  
\usepackage{epsfig} 
\usepackage{subfigure}
\usepackage{lineno}  
\usepackage{amsmath,amssymb}
\usepackage{comment}
\usepackage{multirow}
\usepackage{dcolumn}% Align table columns on decimal point
\usepackage{bm}% bold math
\usepackage{setspace}
\usepackage{multibib}
\includecomment{printrobustnotes}
\includecomment{printallnotes}
\usepackage{xspace}
\usepackage{color}
\def\kforty{$\rm ^{40}K$~}

\newcommand{\gevcc}[1]  {\ensuremath{#1~\mathrm{GeV}/{\rm c}^{2}}}

\begin{document}

\title{An experiment to search for dark matter interactions using sodium iodide detectors}

\author{Govinda~Adhikari}
\affiliation{Department of Physics, Sejong University, Seoul 05006, Republic of Korea}
\author{Pushparaj~Adhikari}
\affiliation{Department of Physics, Sejong University, Seoul 05006, Republic of Korea}
\author{Estella~Barbosa de Souza}
\affiliation{Department of Physics, Yale University, New Haven, CT 06520, USA}
\author{Nelson~Carlin}
\affiliation{Physics Institute, University of S\~{a}o Paulo, 05508-090, S\~{a}o Paulo, Brazil}
\author{Seonho~Choi}
\affiliation{Department of Physics and Astronomy, Seoul National University, Seoul 08826, Republic of Korea} 
\author{Mitra~Djamal}
\affiliation{Department of Physics, Bandung Institute of Technology, Bandung 40132, Indonesia}
\author{Anthony~C.~Ezeribe}
\affiliation{Department of Physics and Astronomy, University of Sheffield, Sheffield S3 7RH, United Kingdom}
\author{Chang~Hyon~Ha}
\email{changhyon.ha@gmail.com}
\affiliation{Center for Underground Physics, Institute for Basic Science (IBS), Daejeon 34126, Republic of Korea}
\author{Insik~Hahn}
\affiliation{Department of Science Education, Ewha Womans University, Seoul 03760, Republic of Korea} 
\author{Antonia~J.F.~Hubbard}
\thanks{Department of Physics and Astronomy, Northwestern University, Evanston, IL 60208, USA}
\affiliation{Department of Physics, Yale University, New Haven, CT 06520, USA}
\author{Eunju~Jeon}
\affiliation{Center for Underground Physics, Institute for Basic Science (IBS), Daejeon 34126, Republic of Korea}
\author{Jay~Hyun~Jo}
\affiliation{Department of Physics, Yale University, New Haven, CT 06520, USA}
\author{Hanwool~Joo}
\affiliation{Department of Physics and Astronomy, Seoul National University, Seoul 08826, Republic of Korea}
\author{Woon~Gu~Kang}
\affiliation{Center for Underground Physics, Institute for Basic Science (IBS), Daejeon 34126, Republic of Korea}
\author{Woosik~Kang}
\affiliation{Department of Physics, Sungkyunkwan University, Suwon 16419, Republic of Korea}
\author{Matthew~Kauer}
\affiliation{Department of Physics and Wisconsin IceCube Particle Astrophysics Center, University of Wisconsin-Madison, Madison, WI 53706, USA}
\author{Bonghee~Kim}
\affiliation{Center for Underground Physics, Institute for Basic Science (IBS), Daejeon 34126, Republic of Korea}
\author{Hyounggyu~Kim}
\affiliation{Center for Underground Physics, Institute for Basic Science (IBS), Daejeon 34126, Republic of Korea}
\author{Hongjoo~Kim}
\affiliation{Department of Physics, Kyungpook National University, Daegu 41566, Republic of Korea}
\author{Kyungwon~Kim}
\affiliation{Center for Underground Physics, Institute for Basic Science (IBS), Daejeon 34126, Republic of Korea}
\author{Nam~Young~Kim}
\affiliation{Center for Underground Physics, Institute for Basic Science (IBS), Daejeon 34126, Republic of Korea}
\author{Sun~Kee~Kim}
\affiliation{Department of Physics and Astronomy, Seoul National University, Seoul 08826, Republic of Korea}
\author{Yeongduk~Kim}
\affiliation{Center for Underground Physics, Institute for Basic Science (IBS), Daejeon 34126, Republic of Korea}
\affiliation{Department of Physics, Sejong University, Seoul 05006, Republic of Korea}
\author{Yong-Hamb~Kim}
\affiliation{Center for Underground Physics, Institute for Basic Science (IBS), Daejeon 34126, Republic of Korea}
\affiliation{Korea Research Institute of Standards and Science, Daejeon 34113, Republic of Korea}
\author{Young~Ju~Ko}
\affiliation{Center for Underground Physics, Institute for Basic Science (IBS), Daejeon 34126, Republic of Korea}
\author{Vitaly~A.~Kudryavtsev}
\affiliation{Department of Physics and Astronomy, University of Sheffield, Sheffield S3 7RH, United Kingdom}
\author{Hyun~Su~Lee}
\email{hyunsulee@ibs.re.kr}
\affiliation{Center for Underground Physics, Institute for Basic Science (IBS), Daejeon 34126, Republic of Korea}
\author{Jaison~Lee}
\affiliation{Center for Underground Physics, Institute for Basic Science (IBS), Daejeon 34126, Republic of Korea}
\author{Jooyoung~Lee}
\affiliation{Department of Physics, Kyungpook National University, Daegu 41566, Republic of Korea}
\author{Moo~Hyun~Lee}
\affiliation{Center for Underground Physics, Institute for Basic Science (IBS), Daejeon 34126, Republic of Korea}
\author{Douglas~S.~Leonard}
\affiliation{Center for Underground Physics, Institute for Basic Science (IBS), Daejeon 34126, Republic of Korea}
\author{Warren~A.~Lynch}
\affiliation{Department of Physics and Astronomy, University of Sheffield, Sheffield S3 7RH, United Kingdom}
\author{Reina~H.~Maruyama}
\affiliation{Department of Physics, Yale University, New Haven, CT 06520, USA}
\author{Frederic~Mouton}
\affiliation{Department of Physics and Astronomy, University of Sheffield, Sheffield S3 7RH, United Kingdom}
\author{Stephen~L.~Olsen}
\affiliation{Center for Underground Physics, Institute for Basic Science (IBS), Daejeon 34126, Republic of Korea}
\author{Byungju~Park}
\affiliation{IBS School, University of Science and Technology (UST), Daejeon 34113, Republic of Korea}
\author{Hyang~Kyu~Park}
\affiliation{Department of Accelerator Science, Korea University, Sejong 30019, Republic of Korea}
\author{Hyeonseo~Park}
\affiliation{Korea Research Institute of Standards and Science, Daejeon 34113, Republic of Korea}
\author{Jungsic~Park}
\thanks{High Energy Accelerator Research Organization (KEK), Ibaraki 319-1106, Japan}
\affiliation{Center for Underground Physics, Institute for Basic Science (IBS), Daejeon 34126, Republic of Korea}
\author{Kangsoon~Park}
\affiliation{Center for Underground Physics, Institute for Basic Science (IBS), Daejeon 34126, Republic of Korea}
\author{Walter~C.~Pettus}
\thanks{Center for Experimental Nuclear Physics and Astrophysics and Department of Physics, University of Washington, Seattle, WA 98195, USA}
\affiliation{Department of Physics, Yale University, New Haven, CT 06520, USA}
\author{Hafizh~Prihtiadi}
\affiliation{Department of Physics, Bandung Institute of Technology, Bandung 40132, Indonesia}
\author{Sejin~Ra}
\affiliation{Center for Underground Physics, Institute for Basic Science (IBS), Daejeon 34126, Republic of Korea}
\author{Carsten~Rott}
\affiliation{Department of Physics, Sungkyunkwan University, Suwon 16419, Republic of Korea}
\author{Andrew~Scarff}
\thanks{Department of Physics and Astronomy, University of British Columbia, Vancouver, BC V6T 1Z1, Canada}
\affiliation{Department of Physics and Astronomy, University of Sheffield, Sheffield S3 7RH, United Kingdom}
\author{Keon~Ah~Shin}
\affiliation{Center for Underground Physics, Institute for Basic Science (IBS), Daejeon 34126, Republic of Korea}
\author{Neil~J.C.~Spooner}
\affiliation{Department of Physics and Astronomy, University of Sheffield, Sheffield S3 7RH, United Kingdom}
\author{William~G.~Thompson}
\affiliation{Department of Physics, Yale University, New Haven, CT 06520, USA}
\author{Liang~Yang}
\affiliation{Department of Physics, University of Illinois at Urbana-Champaign, Urbana, IL 61801, USA}
\author{Seok~Hyun~Yong}
\affiliation{Center for Underground Physics, Institute for Basic Science (IBS), Daejeon 34126, Republic of Korea}

\collaboration{The COSINE-100 collaboration}

\maketitle

%\doublespacing

{\bf 
		Observations of galaxies and primordial radiation suggest that the Universe is made mostly of non-luminous dark matter~\cite{Clowe:2006eq, Ade:2015xua}.  Several types of new fundamental particles have been proposed as candidates for dark matter~\cite{Baer:2014eja} such as weakly interacting massive particles~(WIMPs)~\cite{PhysRevLett.39.165,Goodman:1984dc}, but no definitive signal has been seen despite concerted efforts by many collaborations~\cite{Battaglieri:2017aum}. One exception is the much-debated claim by the DAMA collaboration of a statistically significant annual modulation in the event rate of their experiment~\cite{BERNABEI1998195,Bernabei:2013xsa, Bernabei:2018yyw} with a period and phase consistent with that expected from WIMP dark matter~\cite{Savage:2008er,Baum:2018ekm, Kang:2018qvz}. 
    Several groups have been working to develop experiments with the aim of reproducing DAMA's results using the same target medium~\cite{deSouza:2016fxg,Amare:2015qcn,Fushimi:2015sew,sabre,Adhikari:2015rba}. 
		Here we report results from the initial operation of the COSINE-100 experiment~\cite{Adhikari:2017esn,cosinebg}. COSINE-100 uses sodium iodide as the target medium -- the same medium as DAMA -- and is designed to carry out a model-independent test of DAMA's claim. Initial data based on the first 59.5 days indicate that there is no excess of events over the expected background, confirming that DAMA's annual modulation signal is in severe tension with results from other experiments under the assumption of dark matter having spin independent interactions and the Standard Halo Model~\cite{Tanabashi2018, PhysRevD.33.3495,freese1987, Lewin:1995rx}.  COSINE-100 is now taking data to study the presence of dark matter-induced annual modulation in the event rate of the sodium iodide detectors.
}

COSINE-100 is located at the Yangyang Underground Laboratory in South Korea and began data taking in 2016. 
The experiment utilizes eight low-background thallium-doped sodium iodide crystals arranged in a 4$\times$2 array, giving a total target mass of 106\,kg.  Each crystal is coupled to two photomultiplier tubes~(PMTs) to measure the amount of energy deposited in the crystal.  The sodium iodide crystal assemblies are immersed in 2,200\,L of liquid scintillator, which allows for the identification and subsequent reduction of radioactive backgrounds observed by the crystals~\cite{Park:2017jvs}.  The liquid scintillator is surrounded by copper, lead, and plastic scintillator to reduce the background contribution from external radiation as well as cosmic-ray muons~\cite{Prihtiadi:2017inr}~(Extended Data Fig.~\ref{fig:cosine}).

The data used in this analysis were acquired between 20 October 2016 and 19 December 2016 for a total exposure of 59.5 live days. 
During this two-month period, no significant environmental abnormality or unstable detector performance were observed. 
The analysis was performed with all eight crystals. 
Six of the crystals have light yields of about 15~photoelectrons/keV with analysis threshold of 2\,keV.
The other two crystals have lower light yields and require higher analysis thresholds
of 4\,keV and 8\,keV respectively~\cite{Adhikari:2017esn}. Since their direct impact on the search is not substantial, we discuss the spectra of only six crystals in this Letter. 
When both PMTs on the same crystal register signals that are consistent with
at least one photoelectron within 200\,ns, that crystal is considered to have registered a ``hit.'' 
The outputs of all of the detector elements during 8\,$\mu$s time windows surrounding the hit time are recorded.

A nucleus recoiling from a WIMP interaction is expected to produce a hit in a single crystal.
A set of candidate events are selected by applying several criteria to reject backgrounds. 
Boosted Decision Trees~(BDTs)~\cite{BDT}, i.e. multivariate machine learning
algorithms, are used to characterize the pulse-shapes to discriminate PMT-induced noise events from radiation-induced events. 
Events that had hits in multiple crystals, liquid scintillator, or the muon detector are also 
rejected as multiple hit events. 
Although multiple-hit events are not used for the WIMP search,
they are used to develop the event selection criteria, determine efficiencies, and
model backgrounds.  

Multiple-hit events recorded during the two week calibration campaign with the $^{60}$Co source
provided a large sample of Compton scattering events where a $\gamma$-ray from the $^{60}$Co source scatters from an electron
in one crystal and is detected in another crystal. 
The BDTs are trained for each detector using the multiple-hit events of the $^{60}$Co calibration data, weighted to match the energy spectrum of the expected background, and physics data for both signals and the PMT-induced noise (see Methods). 
The efficiencies of the selection requirements are first measured with the multiple-hit events from the $^{60}$Co source as shown in Fig.~\ref{eff}.

Multiple-hit events from \kforty decay are produced when a 3\,keV X-ray registers in one crystal and its accompanying
1460\,keV $\gamma$-ray in another~\cite{Adhikari:2015rba}. These occur
throughout the data exposure time and provide independent, real-time energy calibrations and efficiency measurements
in the 2--6\,keV region of interest for the WIMP search. 
The efficiencies measured with the multiple-hit events that occur during
the dark matter search exposure, including tagged 3\,keV X-rays from the $^{40}$K, are in agreement with the measured efficiencies using the $^{60}$Co data.
A specialized apparatus that has a monoenergetic 2.42\,MeV neutron beam is used to measure the selection efficiencies of nuclear recoil events. 
This measurement was performed with a small test crystal that was cut from the same ingot as a crystal used for the COSINE-100 experiment. 
The efficiencies determined from the different methods are mutually consistent within a 5\% level of
uncertainty as shown in Fig.~\ref{eff}. The efficiency uncertainties are included as a systematic error.

The remaining dark matter search dataset predominantly originate from environmental $\gamma$ and $\beta$ radiations produced from
the crystals themselves or the nearby surrounding materials. Sources include radioactive contaminants internal to the crystals or on their surfaces, external detector components, and cosmogenic activation~\cite{cosinebg}. 
The background spectrum for each individual crystal is modeled using simulations based on the Geant4
toolkit~\cite{Agostinelli:2002hh}.  Multiple-hit events with measured energies between 2~and~2,000\,keV and single-hit events between 6~and~2,000\,keV are used in the modeling
as described in detail elsewhere~\cite{cosinebg} (see also Methods).
Single-hit events with energies below 6\,keV are excluded to avoid a bias against dark matter signal events. 
Figure~\ref{syst} shows the summed single-hit event spectrum between 2 and 20\,keV for the six 
crystals
compared with the simulated contributions from various sources.  The data points in the 2--6\,keV region of interest 
are within the error bands of the background model.

Several sources of systematic uncertainties were identified and included in this analysis. The largest uncertainties are those associated
with the efficiency, which include statistical errors in the efficiency determination with the  $^{60}$Co
calibration and systematic errors derived from the independent cross-checks.
Uncertainties in the energy resolution and nonlinear responses of the sodium iodide crystals~\cite{nonprop} affect the shapes
of the background and signal spectra. These are studied using tagged 3\,keV X-rays from internal $^{40}$K and 59.5\,keV
$\gamma$-rays from an external $^{241}$Am source.  Different models for $^{210}$Pb decays~\cite{cosinebg} and variations of
the levels of external Uranium and Thorium decay-chain contaminants 
are also accounted for, as are effects of event rate variations and possible distortions in the shapes of the background
model components~(see Methods). 

We used the simulated data to determine the contributions of dark matter-induced nuclear recoils to the
measured energy spectra.  Samples of WIMP-sodium and WIMP-iodine spin independent scattering events were generated for 18 different
WIMP masses, ranging from \gevcc{5} to \gevcc{10,000} using the standard WIMP halo model with the same parameters that
were used for the WIMP interpretation of the DAMA/LIBRA-phase1 signal~\cite{Savage:2008er}. These events were then processed
through the detector simulation and the output events were subjected to the same selection criteria that were applied
to the data.

To search for evidence of dark matter-induced events, binned, maximum likelihood fits to the measured single-hit
energy spectra between 2~and~20\,keV  were performed for each of the 18 WIMP masses. The Bayesian
Analysis Toolkit~\cite{BAT} was used with Probability Density Functions (PDFs) that were based on shapes of the simulated
WIMP signal spectra and the various components of the background model. Uniform priors were used for the signals
and Gaussian priors for the background, with means and uncertainties for each background component set at the values determined from the model fitted to the
data~\cite{cosinebg}.
The systematic uncertainties are included in the fit as nuisance parameters with Gaussian priors.
To be conservative in the assignment of systematic uncertainties, the maximum allowed distortions of the PDF shapes within their
uncertainties are considered. The possibility of correlated rate and shape uncertainties as well
as the uncorrelated bin by bin statistical uncertainties are also considered~(see Methods).
To calculate the expected 90\% confidence level upper limits on WIMP-nucleon scattering cross sections, we performed 1,000 simulated experiments with the expected backgrounds and no additional dark matter signal.

Data were fit to each of the 18 WIMP masses.
An example of a maximum likelihood fit with a \gevcc{10} WIMP signal is presented in Fig.~\ref{fig:avebestfit} (see also Extended Data Fig.~\ref{fig:bestfitC}). 
The summed event spectrum for the six crystals is shown together with the best-fit result.
For comparison, the expected signal for a \gevcc{10} WIMP with a spin independent cross section of $2.35\times10^{-40}$cm$^2$,
the central value of the DAMA/LIBRA-phase1 signal interpreted as a WIMP-sodium interaction, is overlaid in red. 
No excess of events that could be attributed to Standard Halo WIMP interactions are found in the 18 WIMP masses considered. 
The posterior probabilities of signal were consistent with zero in all cases and 90\% confidence level limits are determined. 
Figure~\ref{results} shows the 3$\sigma$ contours of the allowed WIMP mass and cross section values that are associated with the DAMA/LIBRA-phase1 signal~\cite{Savage:2008er} together with the 90\% confidence level upper limits from the COSINE-100 data.

Despite the strong evidence for its existence, the identity of dark matter remains a mystery. COSINE-100, a new sodium iodine-based experiment designed to resolve the long-standing controversy between DAMA collaboration's claim for detection of dark matter and the null results from many other experiments, is now taking data at the Yangyang Underground Laboratory.  Several years of data will be necessary to 
fully confirm or refute DAMA's annual modulation results. However, the first
59.5 days of background data show that the annual modulation signal observed by DAMA is inconsistent with spin independent interaction between WIMPs and sodium or iodine in the context of the Standard Halo Model.

\acknowledgments
We thank the Korea Hydro and Nuclear Power (KHNP) Company for providing underground laboratory space at Yangyang.
This work is supported by:  the Institute for Basic Science (IBS) under project code IBS-R016-A1 and NRF-2016R1A2B3008343, Republic of Korea;
UIUC campus research board, the Alfred P. Sloan Foundation Fellowship,
NSF Grants No. PHY-1151795, PHY-1457995, DGE-1122492 and DGE-1256259,
WIPAC,
the Wisconsin Alumni Research Foundation,
Yale University and
DOE/NNSA Grant No. DE-FC52-08NA28752, United States; 
STFC Grant ST/N000277/1 and ST/K001337/1, United Kingdom;
and CNPq and Grant No. 2017/02952-0 FAPESP, Brazil.

\section*{Author Information}
{\bf Contributions}
Y.D.K., H.S.L., R.H.M., and N.J.C.S. conceived the COSINE-100 experiment.
Its design and installation were led by K.S.P. and C.H. and carried out by members of the COSINE-100 collaboration.
Operation and maintenance were organized by C.H. with the support from on-site crews, W.G.K., B.H.K., and S.H.Y.
J.L., J.S.P., J.H.J., G.A., P.A., H.P., C.H, W.G.T., E.B.S., H.S.L., and K.W.K. contributed in data acquisition, production, and verification.
H.W.J., H.P., and K.W.K. provided nuclear recoil data.
P.A., G.A., J.S.P., K.W.K., H.P., N.Y.K., and C.H. performed source calibrations.
H.K., N.Y.K., C.H., and H.S.L. developed slow control framework.
J.H.J. and W.G.T. developed data monitoring package.
N.Y.K., J.Y.L., and Y.J.K. provided radiopurity of detector materials.
G.A., J.S.P., and N.Y.K produced the liquid scintillator.
Background simulations were performed by F.M., E.J.J., P.A., W.G.T., and E.B.S.
C.H. and P.A analyzed data and the simulated data.
This manuscript and plots were produced by C.H. and H.S.L., and
edited by R.H.M., S.L.O, N.J.C.S., and other members of the collaboration.
All authors have participated in online shift and approved the manuscript.
Authors are alphabetically listed by their last names.\\

{\bf Competing interests} The authors declare no competing interests.\\

{\bf Correspondence and requests for materials} should be addressed to C.H. and H.S.L. \\

{\bf Fig.1 Event selection efficiency.} 
    A variety of methods are used to evaluate event selection efficiencies.
    The statistical error bands (68\% confidence interval) of the event selection efficiencies determined from the $^{60}$Co calibration data are
    shown as green shaded regions and are compared with efficiencies determined from multiple-hit events~(red diamond),
    internal $^{40}$K coincidence events~(black square), and the nuclear recoil calibration data~(blue circle) for one of the crystals.
    Horizontal error bars depict the data bin width. Vertical error bars are 68\% confidence intervals.

		{\bf Fig.2 Measured and simulated energy spectra.}
    Summed energy spectrum for the six 
    crystals (black filled circles shown with 68\% confidence level error bars) and the expected background (blue
    line) are compared.  Contributions to the background from internal radionuclide contaminations (primarily $^{210}$Pb and $^{40}$K),
    $^{210}$Pb on the surface of the crystals and nearby materials, cosmogenic activation (mostly $^{109}$Cd and $^3$H),
    and external backgrounds~(mostly $^{238}$U and $^{232}$Th) are indicated. The green (yellow) bands are the 68\% (95\%) confidence level intervals for the background model.
		
		{\bf Fig.3 Fit results for a 10\,GeV/c$^{2}$ WIMP mass.}
    The data points (black filled circles shown with 68\% confidence level error bars) show the summed energy spectra from the six 
    crystals and the solid blue line
    shows the result for the \gevcc{10} WIMP mass fit.  The expected signal excess above background for a
    \gevcc{10} WIMP mass and a spin independent WIMP-nucleon cross section of 2.35$\times$10$^{-40}$cm$^{2}$ is shown as
    a solid red line. The green~(yellow) bands are the 68\% (95\%) confidence level intervals
    for the background model.
    The lower panel shows the data-best fit residuals normalized to the best fit.
    Here the bands of systematic uncertainty and the expected DAMA/LIBRA-phase1 signal spectrum are shown. 

    {\bf Fig.4 Exclusion limits on the WIMP-nucleon spin independent cross section.}
    The observed (filled circles with black solid line) 90\% confidence level exclusion limits on the WIMP-nucleon
    spin independent cross section from the first 59.5~days data of COSINE-100 are shown together with the
    65\% and 95\% probability bands for the expected 90\% confidence level limit
    assuming the background-only hypothesis. 
    The limits are compared with a WIMP interpretation of DAMA/LIBRA-phase1 of 3$\sigma$ allowed region for 
		the WIMP-sodium~(red-dot-contour) and the WIMP-iodine~(blue-dot-contour) scattering hypothesis~\cite{Savage:2008er}. The limits from NAIAD~\cite{Alner:2005kt}, the only other sodium iodide based experiment to set a competitive limit, are shown in magenta.

\clearpage

\ifx \standalonesupplemental\undefined
\setcounter{page}{1}
\setcounter{figure}{0}
\setcounter{table}{0}
\renewcommand{\figurename}{Fig.}
\renewcommand{\tablename}{Table}

\begin{figure}[!htb]
  \begin{center}
    \includegraphics[width=0.8\textwidth]{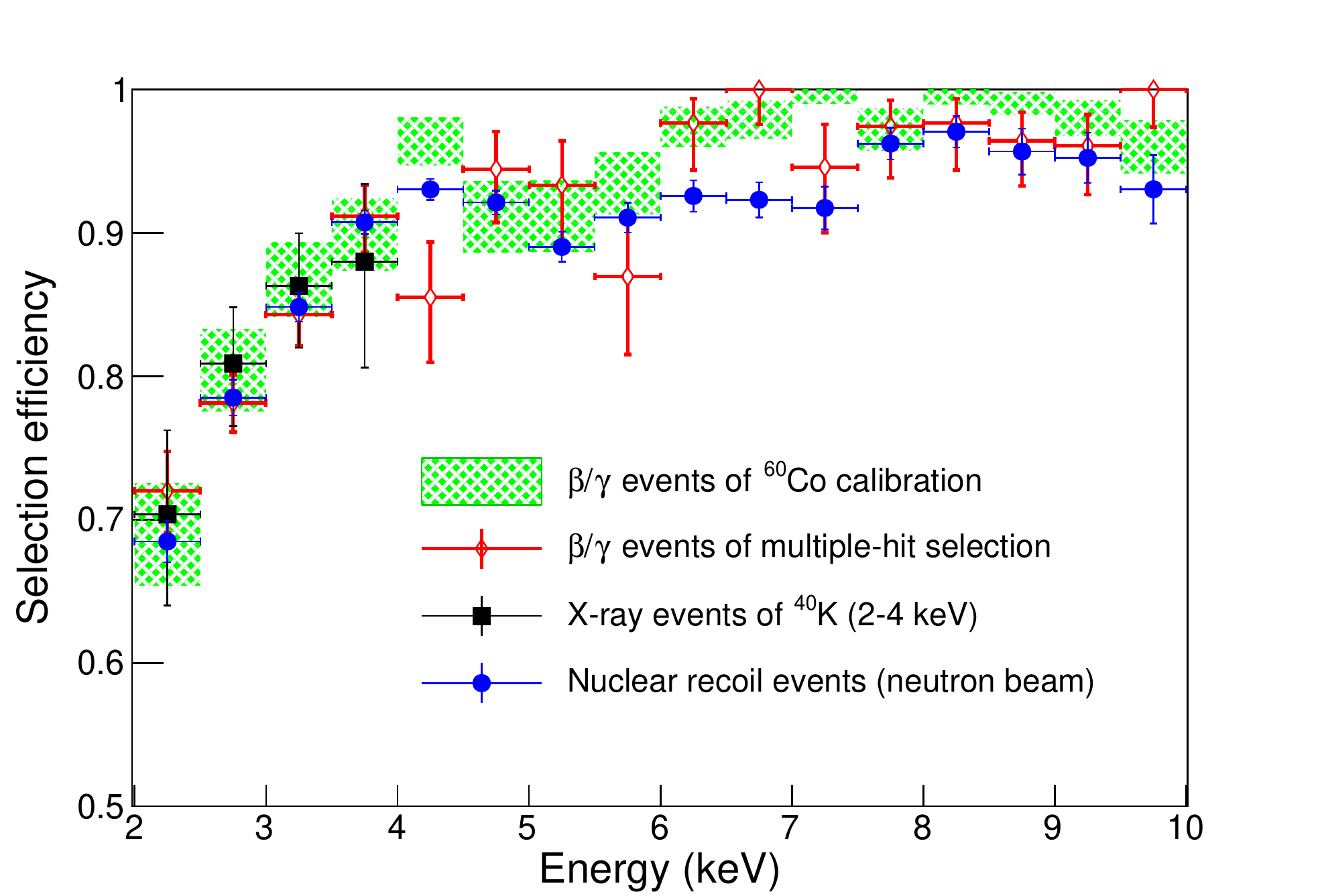}
  \end{center}
  \caption{
    {\bf Event selection efficiency.} 
    A variety of methods are used to evaluate event selection efficiencies.
    The statistical error bands (68\% confidence interval) of the event selection efficiencies determined from the $^{60}$Co calibration data are
    shown as green shaded regions and are compared with efficiencies determined from multiple-hit events~(red diamond),
    internal $^{40}$K coincidence events~(black square), and the nuclear recoil calibration data~(blue circle) for a small test crystal.
    Horizontal error bars depict the data bin width. Vertical error bars are 68\% confidence intervals.
  }
  \label{eff}
\end{figure}

\begin{figure}[!htb]
  \begin{center}
    \includegraphics[width=0.8\textwidth]{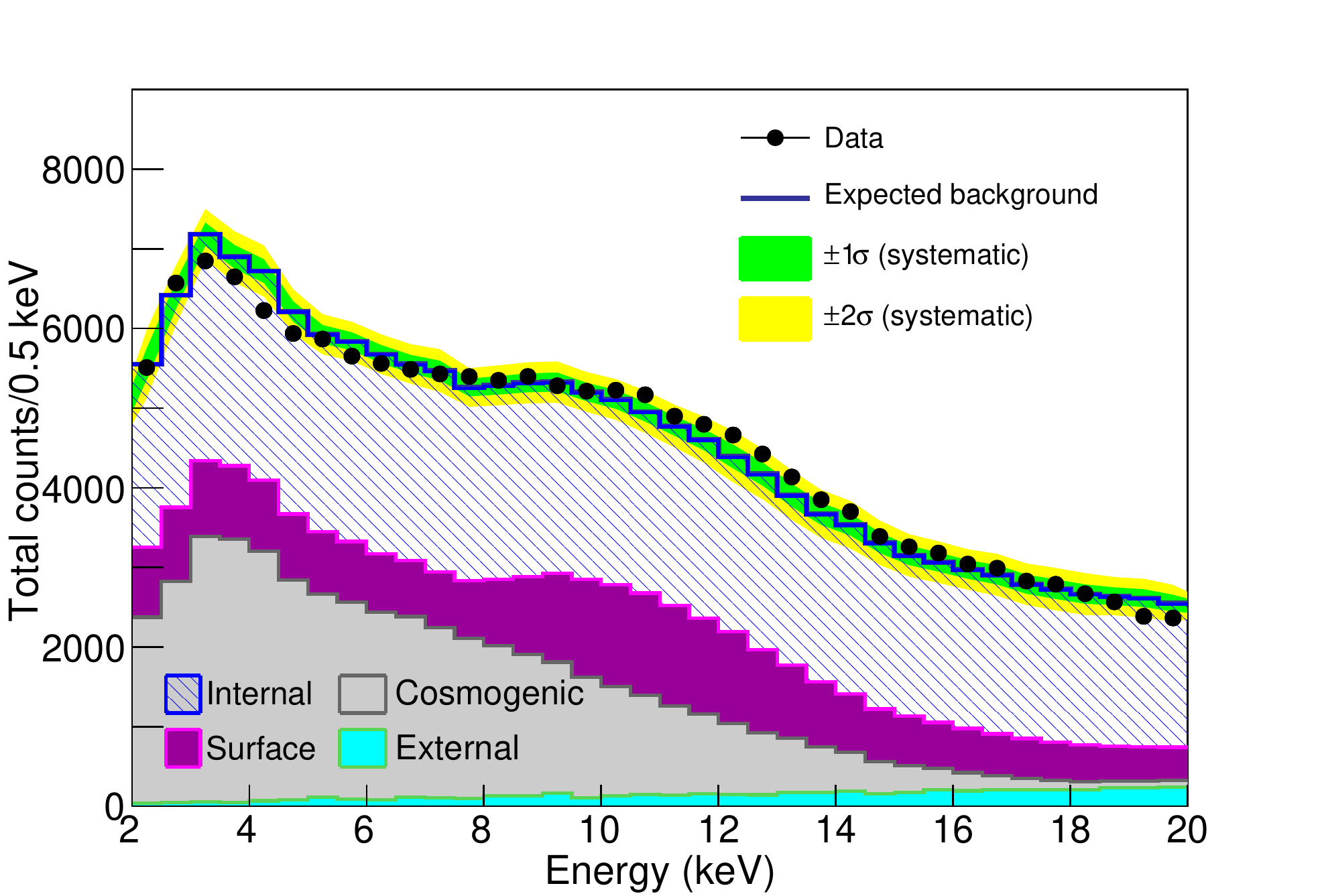} 
  \end{center}
  \caption{ 
    {\bf Measured and simulated energy spectra.}
    Summed energy spectrum for the six 
    crystals (black filled circles shown with 68\% confidence level error bars) and the expected background (blue
    line) are compared.  Contributions to the background from internal radionuclide contaminations (primarily $^{210}$Pb and $^{40}$K),
    $^{210}$Pb on the surface of the crystals and nearby materials, cosmogenic activation (mostly $^{109}$Cd and $^3$H),
    and external backgrounds~(mostly $^{238}$U and $^{232}$Th) are indicated. The green (yellow) bands are the 68\% (95\%) confidence level intervals for the background model.
  }
  \label{syst}
\end{figure}

\begin{figure}[!htb]
  \includegraphics[width=0.8\textwidth]{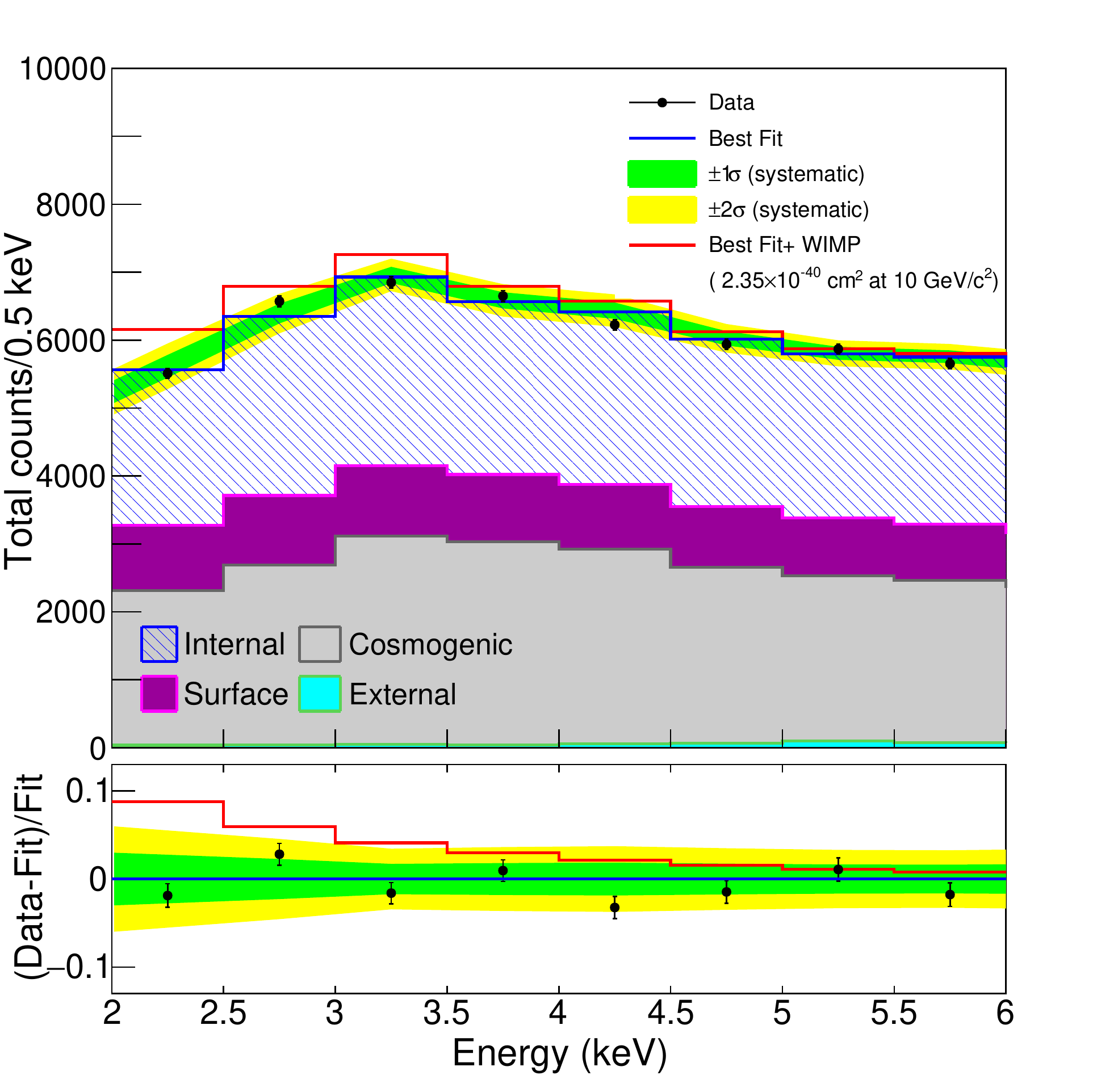}
  \caption{
    {\bf Fit results for a 10\,GeV/c$^{2}$ WIMP mass.}
    The data points (black filled circles shown with 68\% confidence level error bars) show the summed energy spectra from the six 
    crystals and the solid blue line
    shows the result for the \gevcc{10} WIMP mass fit.  The expected signal excess above background for a
    \gevcc{10} WIMP mass and a spin independent WIMP-nucleon cross section of 2.35$\times$10$^{-40}$cm$^{2}$ is shown as
    a solid red line. The green~(yellow) bands are the 68\% (95\%) confidence level intervals
    for the background model.
    The lower panel shows the data-best fit residuals normalized to the best fit.
    Here the bands of systematic uncertainty and the expected DAMA/LIBRA-phase1 signal spectrum are shown. 
  }
\label{fig:avebestfit}
\end{figure}

\begin{figure}[!htb]
  \begin{center}
    \includegraphics[width=0.8\textwidth]{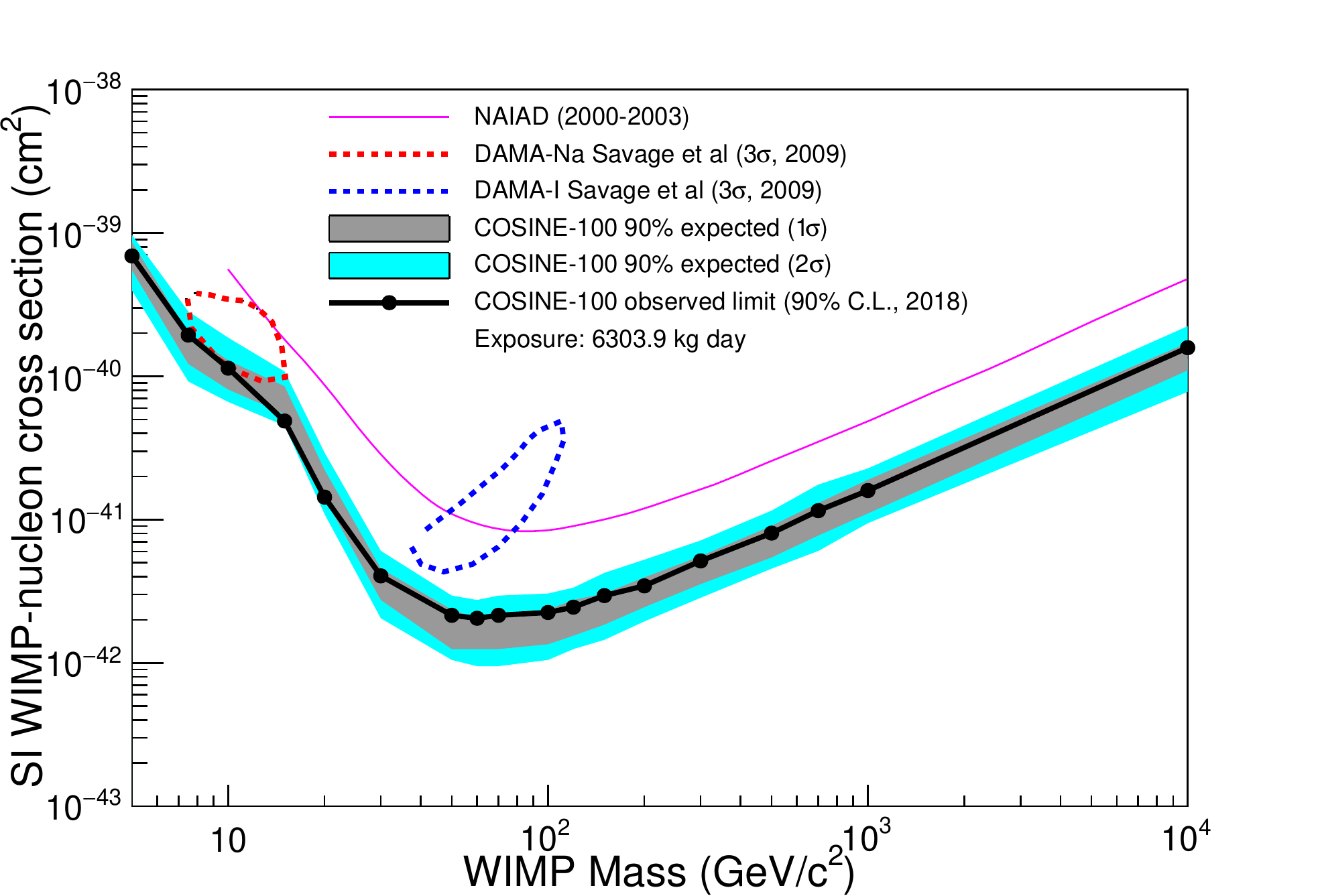}
  \end{center}
  \caption{
    {\bf Exclusion limits on the WIMP-nucleon spin independent cross section.}
    The observed (filled circles with black solid line) 90\% confidence level exclusion limits on the WIMP-nucleon spin independent cross section from the first 59.5~days data of COSINE-100 are shown together with the 65\% and 95\% probability bands for the expected 90\% confidence level limit assuming the background-only hypothesis.  The limits are compared with a WIMP interpretation of DAMA/LIBRA-phase1 of 3$\sigma$ allowed region for the WIMP-sodium~(red-dot-contour) and the WIMP-iodine~(blue-dot-contour) scattering hypothesis~\cite{Savage:2008er}. The limits from NAIAD~\cite{Alner:2005kt}, the only other sodium iodide based experiment to set a competitive limit, are shown in magenta. 
}
\label{results}
\end{figure}

\clearpage

\ifx \standalonesupplemental\undefined
\setcounter{page}{1}
\setcounter{figure}{0}

\section*{Methods}

The COSINE-100 experiment is located 700\,m below the surface at the Yangyang Underground
Laboratory in eastern Korea.
A cut-out view of the detector is shown in Extended Data Fig.~\ref{fig:cosine}.
It is comprised of an array of eight sodium iodide ``NaI(Tl)'' scintillating crystals (total mass 106\,kg)
immersed in 2,200\,L of liquid scintillator contained in an acrylic box that is surrounded by copper and lead shielding.
Plastic scintillators surround the entire apparatus to detect cosmic-ray muons that penetrate the apparatus. 
External radiation is attenuated
by the lead, copper and liquid scintillator shields and signals
from the
liquid scintillator and muon detectors are used to identify background events that are induced by radiation sources
in or near the crystals and cosmic-ray muons.
More details about the experimental site, including the fluxes of
cosmic-ray muons and neutrons, and data acquisition system can be found elsewhere~\cite{Adhikari:2017esn,Prihtiadi:2017inr,Adhikari:2018fpo}.

\subsection*{event selection}

Pulse shapes from the detector are recorded when both PMTs on a crystal record signals that correspond to at least one single photoelectron within 200\,ns. In the offline analysis, events are rejected if they occur
within 30\,ms of a signal from any of the surrounding muon detectors or if there is a signal in the
liquid scintillator within 4\,$\mu$s.
Events with and without accompanying hit crystals in an $8\mu$s time window are classified as multiple-hit and single-hit, respectively.
Events are further classified according to their energy: 2--70\,keV are low energy and 70--2,000\,keV are high energy.

Extended Data Figure~\ref{fig:waves}~(panel a)) shows an averaged waveform for radiation-induced scintillation light
signals in the NaI(Tl) crystal detectors, where the characteristic 250\,ns NaI(Tl) scintillation light decay time is evident.
In contrast, PMT noise pulses, which are considerably more frequent, decay faster, with decay times ranging between
20\,ns and 50\,ns, as shown in panel b).  Some detectors intermittently produce
events that have both slow-rise and decay-times as shown in the figure's panel c). These are attributed to PMT discharges.

Boosted Decision Trees (BDTs) are used to separate signal from noise events. The fast PMT noise-induced events are efficiently
removed by a BDT that is based on the amplitude-weighted average time of a signal,
the ratios of the leading-edge and trailing-edge charge sums relative to total charge, 
and the balance of deposited energies between the two PMTs. 
This is trained with a sample of signal-rich, energy-weighted,
multiple-hit events from the $^{60}$Co calibration campaign, and single-hit events in the WIMP-search data sample, which are mostly triggered by PMT noise.
A second boosted decision tree (BDTA) that includes
weighted higher-order time moments is effective at eliminating discharge events.  
Extended Data Figure~\ref{fig:bdt} shows two-dimensional scatter plots of the BDT vs BDTA outputs for two separate crystals,
one with and the other without PMT discharge signals.
Events that are above and to the right of the dashed red lines in the figure are retained.

\subsection*{background modeling}
The primary background components of the crystal energy spectra are from internal
$^{238}$U, $^{232}$Th, $^{40}$K and $^{210}$Pb contaminations in the bulk material of the crystal,
plus additional $^{210}$Pb on the surfaces of the crystal and its reflecting wrapping foil, caused
by exposure to atmospheric Radon during the crystals' encapsulation~\cite{cosinebg}.
In addition, we considered
background from external sources such as $^{238}$U, $^{232}$Th and $^{40}$K contaminations in the PMTs,
liquid scintillator, and the bulk material of the surrounding shields. The modeling of these contributions
used starting values that were based on radioassay results from an underground,
high-purity Ge detector~\cite{Sala:2016wlz}. The modeling of contributions from cosmogenic activity in
the crystals was guided by measured surface production rates in NaI(Tl)~\cite{pettus} and the above-
and below-ground histories of each individual crystal. 

In 10.7\% of $^{40}$K decays, a $\sim$3\,keV K-shell X-ray (or Auger electron) is produced in
coincidence with a 1460\,keV $\gamma$-ray.  Since this results in a peaking background in the WIMP
search region of interest (ROI), it is of particular concern. However, in the COSINE-100 detector,
about 80\% of these 3\,keV X-rays are tagged by the detection of its accompanying 1460\,keV $\gamma$-ray
in one of the other crystals or in the liquid scintillator and, thus, can be vetoed.  The measured rate for these
tagged events is used to establish the contribution of untagged 3\,keV $^{40}$K-induced events to the
background in the single-hit spectrum's ROI in each crystal.

Extended Data Figure~\ref{fig:coverage} shows the results of the model fits to the data
for the four categories of events (single-hit and multiple-hit events in low and high energy), with $1\sigma$ and $2\sigma$ uncertainty bands indicated in green
and yellow, respectively~\cite{cosinebg}. All four distributions were fit simultaneously. To avoid
biasing the WIMP search, the 2~to~6\,keV region  of the low-energy, single-hit spectrum (panel a) in the figure) is not included in the fitting. The fit model indicates that the main
contributions in the 2--6\,keV ROI are from internal $^{40}$K and $^{210}$Pb, and cosmogenic $^{109}$Cd
and $^3$H. The $^{109}$Cd contribution was independently confirmed by a time-dependent analysis.

\subsection*{systematic uncertainties}

The analysis results are limited by the systematic uncertainties.  
Errors in the selection efficiency, the energy resolution, the energy scale and 
background modeling technique translate into uncertainties in the shapes of the signal and background
component PDFs that are used in the likelihood fit and affect the results. 
These quantities are allowed to vary within their errors in the likelihood as nuisance parameters. 
Of these, the systematic errors associated with the efficiencies have the largest effects on the results.
Uncertainties in the efficiencies are determined by the statistical errors
from the multiple-hit $^{60}$Co source data efficiency measurements
and their stability has been verified with independent datasets shown in Fig.~\ref{eff}.
The efficiency systematic that maximally covers the statistical errors in the region of interest mimics the shape of a WIMP signal.

For most of the energy range,
the resolutions and scales are well measured with internal radioactive peaks and external calibrations. 
However, since external source measurements are impractical for energies below 10 keV, the resolution
and scale values for these energies are determined with the samples of tagged 3\,keV X-rays from the
internal $^{40}$K contamination. For these, statistical errors dominate and are taken as the systematic
spread from these quantities.  We used changes that occur in the background model when the simulation
is done with different locations of the U/Th contamination in the PMTs, and alternative Geant4 methods for X-ray production of $^{210}$Pb as the systematic error from this source.
The inclusion of the total systematic uncertainties degrades the sensitivity by a factor of 2.3.

\subsection*{WIMP extraction Bayesian fit}
A Bayesian analysis with a likelihood formulated in Eq.~\ref{equ:llh} was performed
and this fitter, more computationally demanding than the background modeling fits, was run
with the WIMP search data (low-energy single-hit spectrum) between 2 and 20 keV.
The function that is maximized has the form
\begin{equation}
  \mathcal{L} = \prod^{N_{ch}}_i\prod^{N_{bin}}_j \frac{\mu^{n_{ij}}_{ij}e^{-\mu_{ij}}}{n_{ij}!}\prod^{N_{bkg}}_ke^{-\frac{(x_k-\alpha_k)^2}{2\sigma_k^2}}\prod^{N_{syst}}_le^{-\frac{x_l^2}{2\sigma_l^2}},
\label{equ:llh}
\end{equation}
where, $N_{ch}$ is the number of crystals,
$N_{bin}$ is the number of bins in each histogram,
$N_{bkg}$ is the number of background components,
$N_{syst}$ is the number of systematic nuisance parameters,
$n_{ij}$ is the number of observed counts and $\mu_{ij}$ is the total model expectation by summing all
$N_{bkg}$ background components and a WIMP signal component after application of a shape change due to $N_{syst}$ systematic effects.
In the first product of Gaussians, $x_k$ is the value of the $k^{th}$ background component, $\alpha_k$ is the mean value and $\sigma_k$
is its 68\% error.
The second product of Gaussians $x_l$ is the $l^{th}$ systematic parameter and $\sigma_l$ is its error.

To avoid biasing the WIMP search, the fitter was developed and tested with simulated event samples.
All eight crystals are fit simultaneously with a common WIMP signal model for each assumed WIMP mass and with fits performed for 18 different WIMP masses between 5 GeV/c$^2$ and 10,000 GeV/c$^2$.
The shapes of the WIMP signal energy spectra are determined from simulations based on the standard WIMP halo model
with parameters taken from Ref.~10.  To relate the simulated WIMP signals, which are caused by nuclear recoils,  
to our energy scale, which is calibrated with electron recoils, we use the same NaI(Tl) quenching factors
that were used in interpretations of the DAMA/LIBRA-phase1 signal, (Q$_{Na}$=0.3 and Q$_I$=0.09)~(Quenching factors are the ratio of scintillation-light energy determinations
for nuclear- and electron recoils of the same energy).

We use Gaussian priors for the normalizations of the background components and
the systematic nuisance parameters for efficiencies, the energy resolutions,  and the energy scales.
The initial values for the background
component normalizations are taken from the above-described fits that do not use the single-hit
events in the 2--6\,keV ROI.
The final fit values for all nuisance parameters are within
$\pm1\sigma$ of their initial values.

\bigskip
\bigskip

{\bf Code availability}
All analysis data are in the ROOT~(https://root.cern.ch) format.
Analysis toolkits such as ROOT including BDT and BAT~(https://bat.mpp.mpg.de) are available online. Our custom codes will be made available to GitHub~(https://github.com) upon reasonable request.

\bigskip

{\bf Data availability}
The data that support the findings of this study are available from the corresponding authors upon reasonable request. 
Source Data for Fig. 1, 2, 3, and 4 are provided with the paper.

\bigskip
\bigskip

{\bf Extended Data Fig.1 The COSINE-100 detector.}
The detector is contained within a nested arrangement of shielding components shown in schematic a), as indicated by different colors.
The main purpose of the shield is to provide 4\,$\pi$ coverage against external radiation from various background sources. 
The shielding components include
plastic scintillator panels (blue), a lead brick enclosure (grey) and a copper box (reddish brown).
The eight encapsulated sodium iodide crystal assemblies (schematic c)) are located inside the copper box
and are immersed in scintillating liquid, as shown in schematic b).

{\bf Extended Data Fig.2 Typical  waveforms from the COSINE-100 photomultiplier tubes
for 2--6 keV signals.}
Panel a) shows the shape of $\beta$/$\gamma$ scintillation signals;
these have a fast rise and then fall-off with a decay time of about 250\,ns.
Panels b) and c) show background waveforms from PMT noise and external discharge, respectively.
The waveform from WIMPs is expected to closely resemble the $\beta$/$\gamma$ waveforms.

{\bf Extended Data Fig.3 The BDT output (horizontal) versus the BDTA output (vertical).}
Events with energies below 10\,keV are shown for two separate crystals.
The events to the right and above the red dotted lines are scintillation events induced by real particle-crystal interactions.
PMT noise events are to the left of the vertical dotted lines in panels a) and b);  PMT discharge events are below the horizontal dotted line in panel a).

{\bf Extended Data Fig.4 A comparison between data and simulation.}
Four categories of data are shown: a) single-hit low-energy between 2--70\,keV;
b) single-hit between 70--3,000\,keV; c) multiple-hit between 2--70\,keV; d) multiple-hit between 70--3,000\,keV. The black points with 68\% CL error bars are data and
the green (yellow) band shows the $\pm1\sigma$ ($\pm2\sigma$) uncertainty range of the model.
The peak near 3\,keV in the multiple-hit, low-energy spectrum (panel c)) is due the tagged $^{40}$K events.
A small inset a--1) in panel a) shows a zoomed-in view in the region of interest after efficiency corrections are applied.
The major contributors to the radioactive background are labelled. 

{\bf Extended Data Fig.5 Crystal-by-crystal fit results.}
The points with 68\% CL error bars show the measured energy spectra for each of the six crystals a)--f). The fit results are shown as blue histograms with the $\pm 1\sigma$ ($\pm 2\sigma$) error bands shown in green (yellow).
To compare the signal strength of the DAMA-Na region with our data, a 10 GeV/c$^2$ WIMP signal at 2.35$\times$10$^{-40}$ cm$^2$ (DAMA-Na central region) is indicated for each crystal as a red histogram.
The fit residuals, together with the expectations for the 10\,GeV/$c^2$ WIMP signal are also shown.

\clearpage

\ifx \standalonesupplemental\undefined
\setcounter{page}{1}
\setcounter{figure}{0}
\setcounter{table}{0}

\renewcommand{\figurename}{Extended Data Fig.}
\renewcommand{\tablename}{Extended Data Table}

\begin{figure*}
\includegraphics[width=1.0\textwidth]{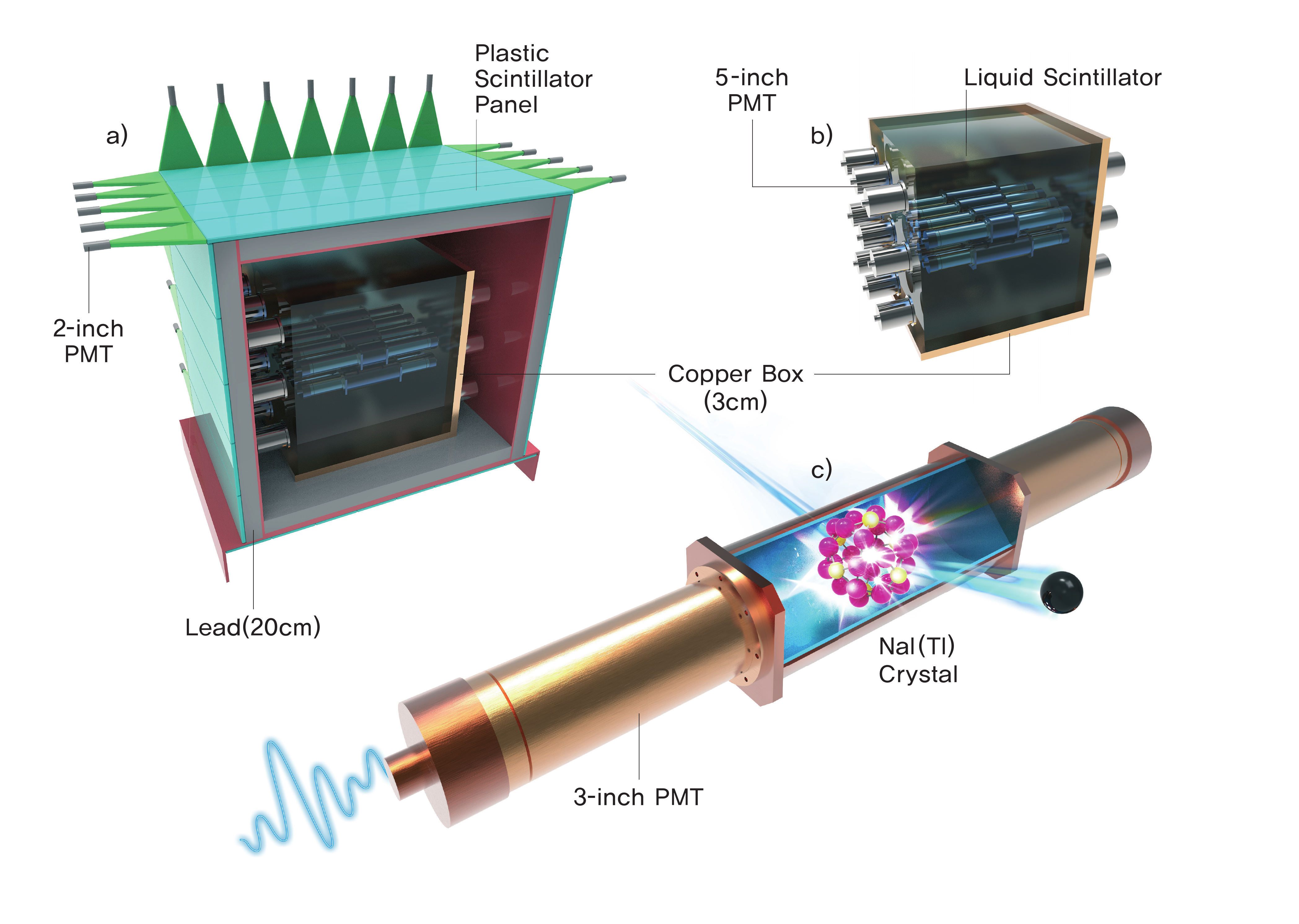}
\caption{
        {\bf The COSINE-100 detector.}
The detector is contained within a nested arrangement of shielding components shown in schematic a), as indicated by different colors.
The main purpose of the shield is to provide 4\,$\pi$ coverage against external radiation from various background sources. 
The shielding components include
plastic scintillator panels (blue), a lead brick enclosure (grey) and a copper box (reddish brown).
The eight encapsulated sodium iodide crystal assemblies (schematic c)) are located inside the copper box
and are immersed in scintillating liquid, as shown in schematic b).
}
\label{fig:cosine}
\end{figure*}

\begin{figure}
\begin{tabular}{ccc}
\includegraphics[width=0.5\textwidth]{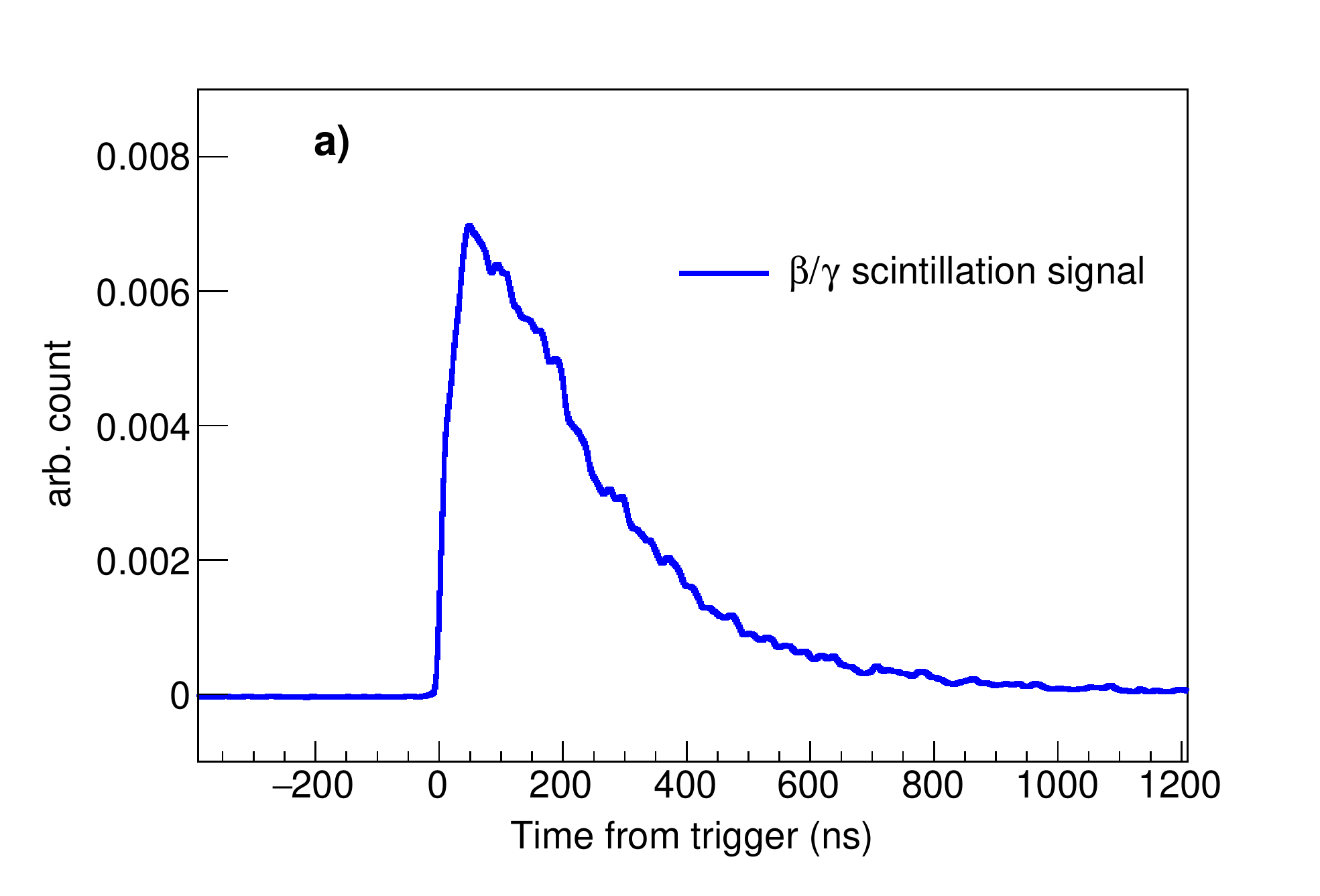} \\
\includegraphics[width=0.5\textwidth]{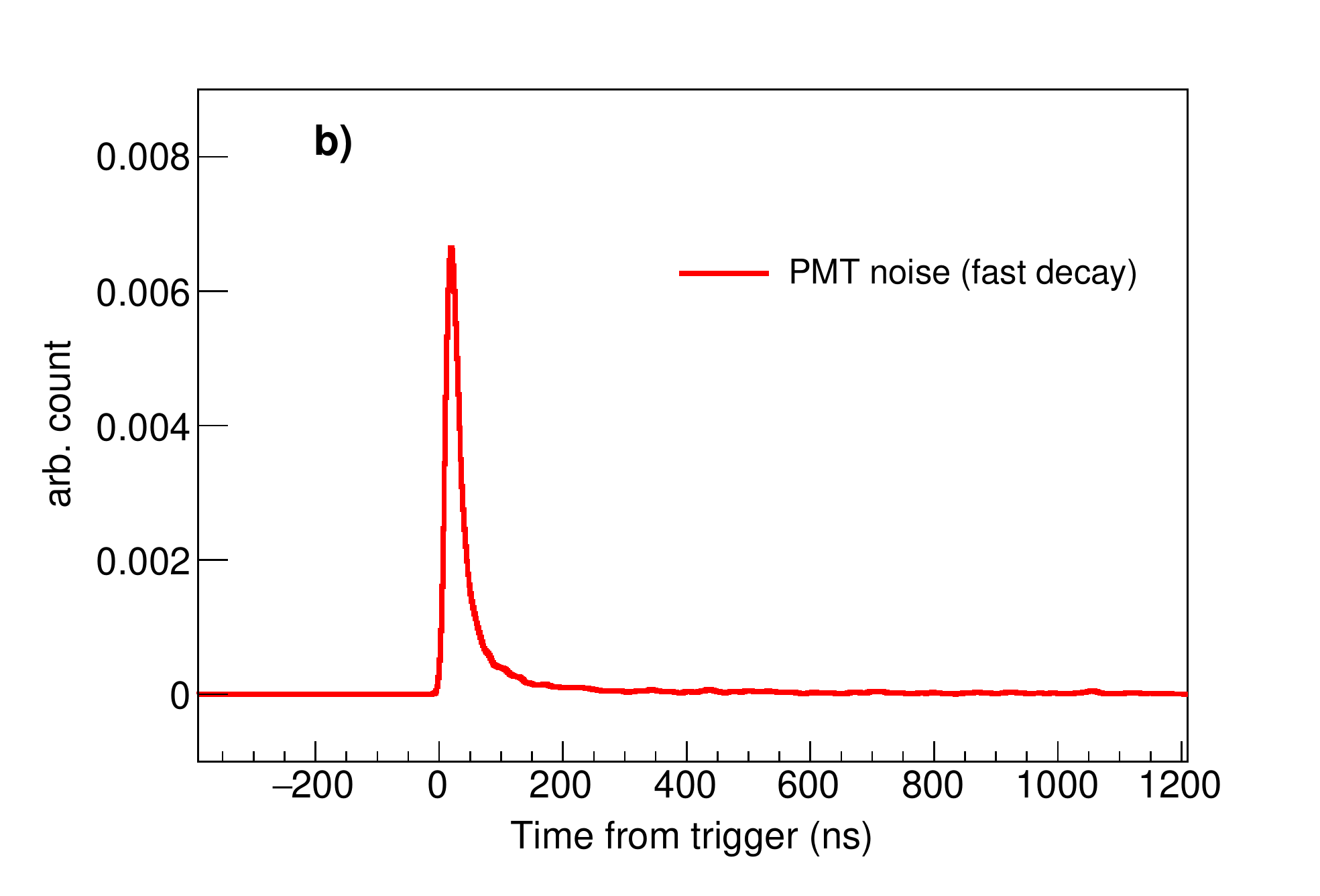} \\
\includegraphics[width=0.5\textwidth]{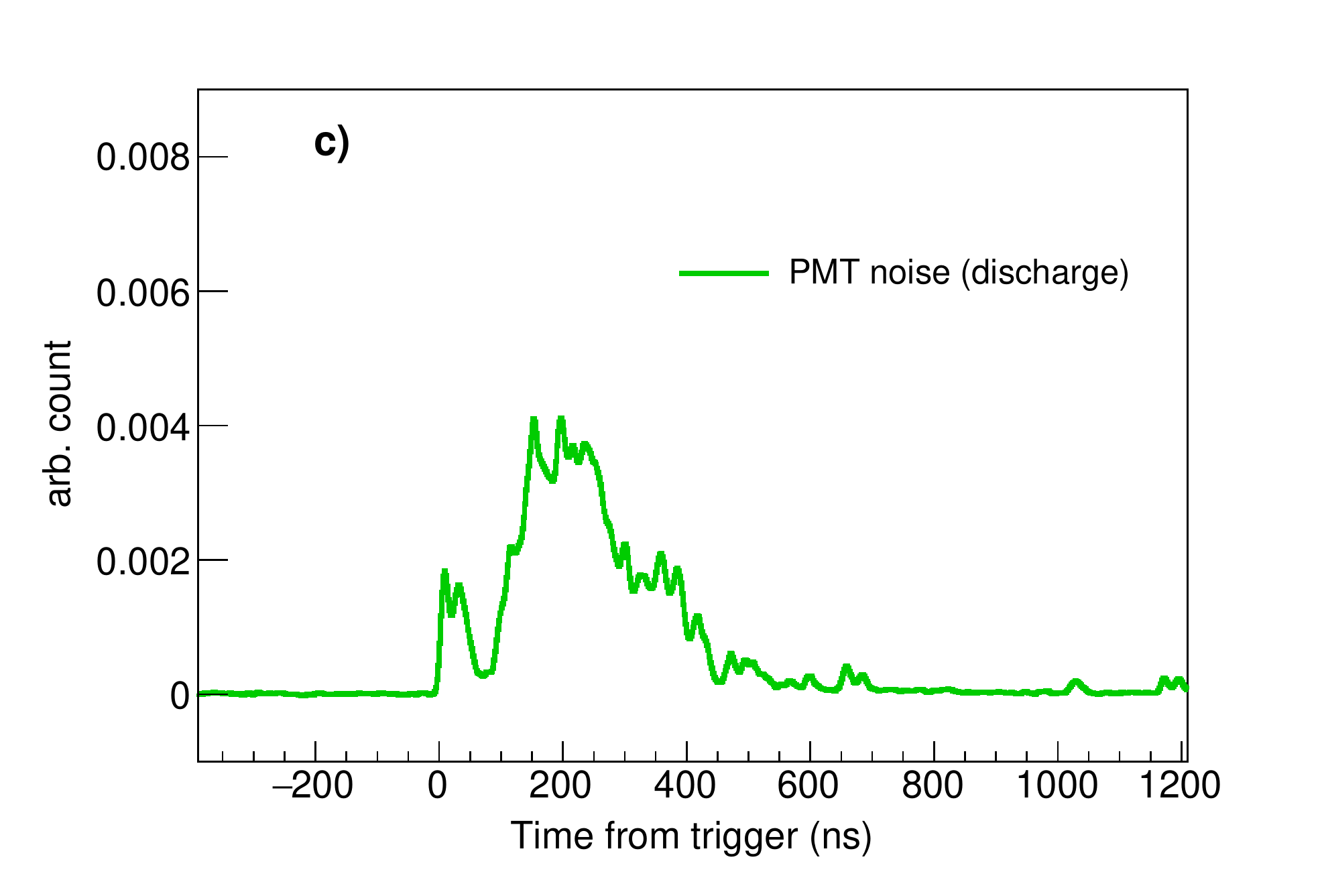} 
\end{tabular}

\caption{
{\bf Typical  waveforms from the COSINE-100 photomultiplier tubes
for 2--6 keV signals.}
Panel a) shows the shape of $\beta$/$\gamma$ scintillation signals;
these have a fast rise and then fall-off with a decay time of about 250\,ns.
Panels b) and c) show background waveforms from PMT noise and external discharge, respectively.
The waveform from WIMPs is expected to closely resemble the $\beta$/$\gamma$ waveforms.
}
\label{fig:waves}
\end{figure}

\begin{figure*}

\begin{tabular}{cc}
\includegraphics[width=0.5\textwidth]{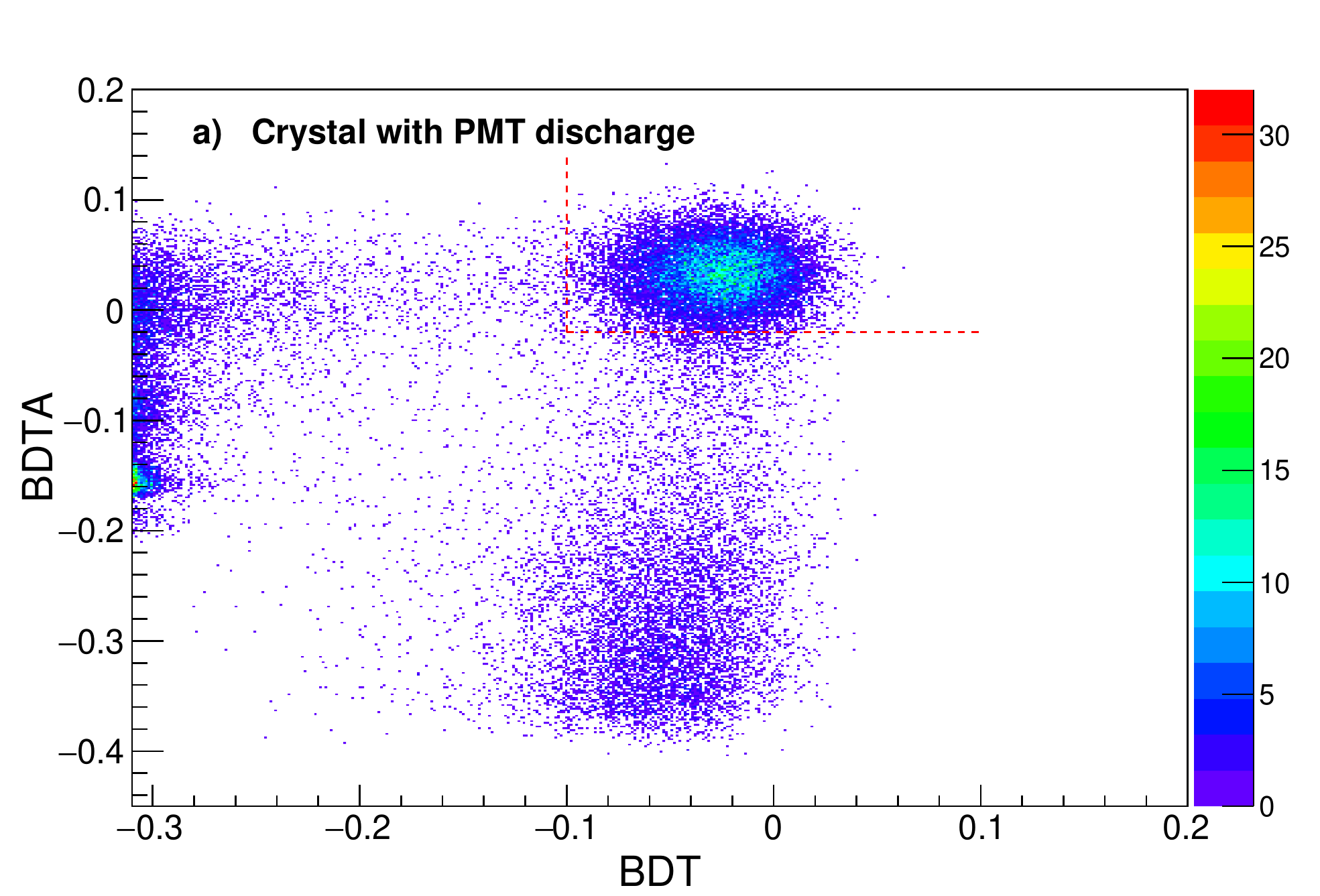}
\includegraphics[width=0.5\textwidth]{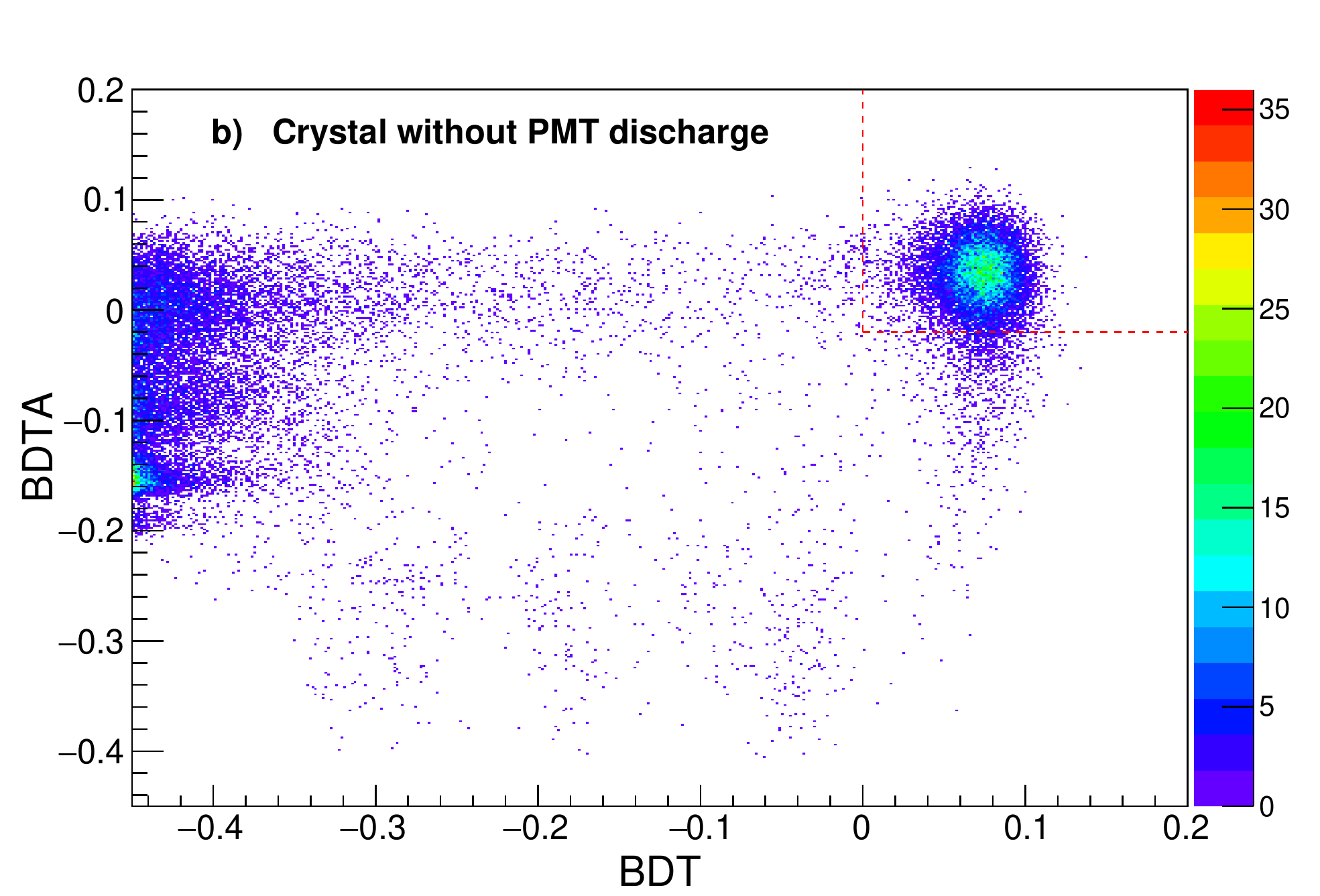}
\end{tabular}
\caption{
{\bf The BDT output (horizontal) versus the BDTA output (vertical).}
Events with energies below 10~keV are shown for two separate crystals.
The events to the right and above the red dotted lines are scintillation events induced by real particle-crystal interactions.
PMT noise events are to the left of the vertical dotted lines in panels a) and b);  PMT discharge events are below the horizontal dotted line in panel a).
}
\label{fig:bdt}
\end{figure*}

\begin{figure*}
\begin{tabular}{cc}
\includegraphics[width=0.5\textwidth]{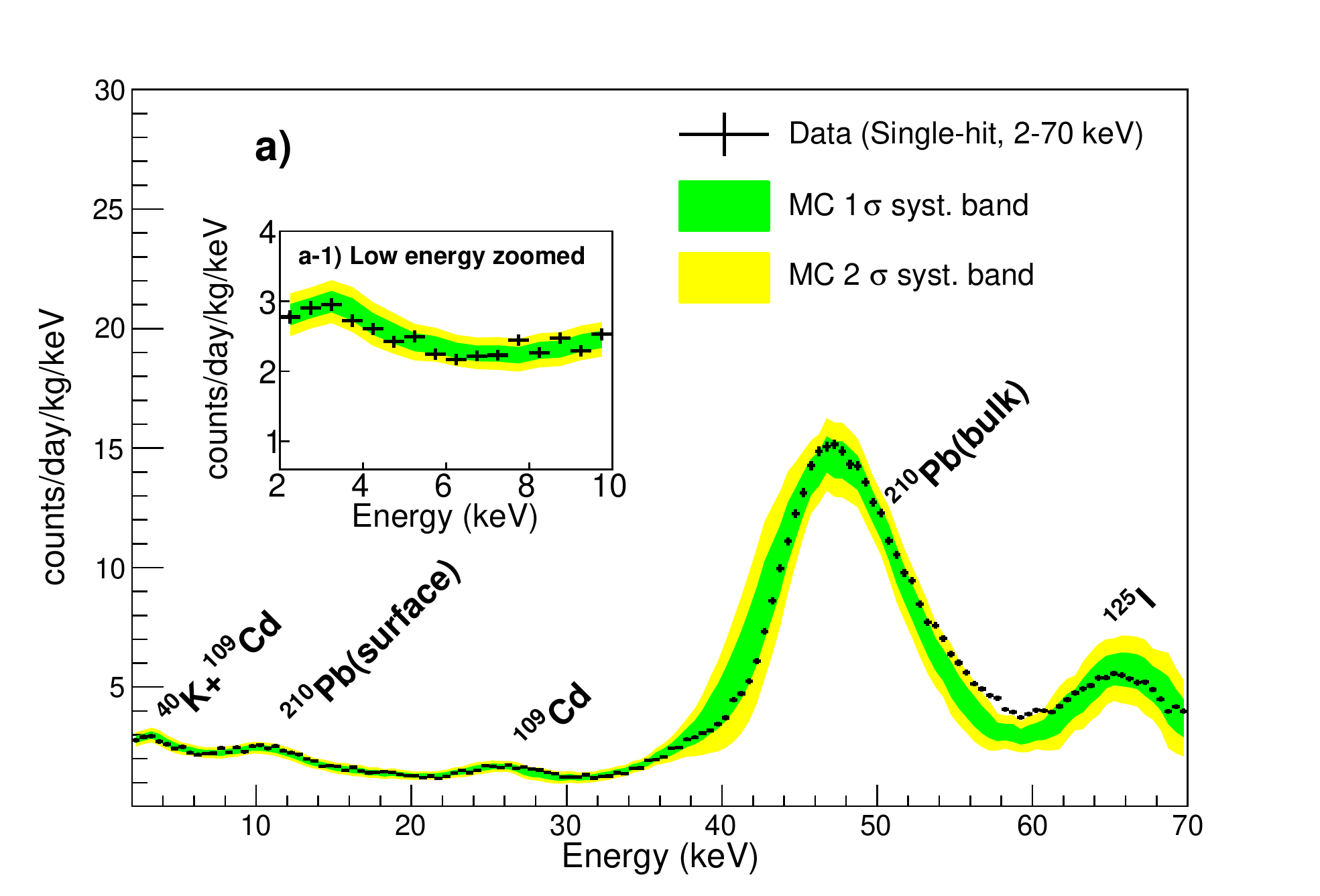}
\includegraphics[width=0.5\textwidth]{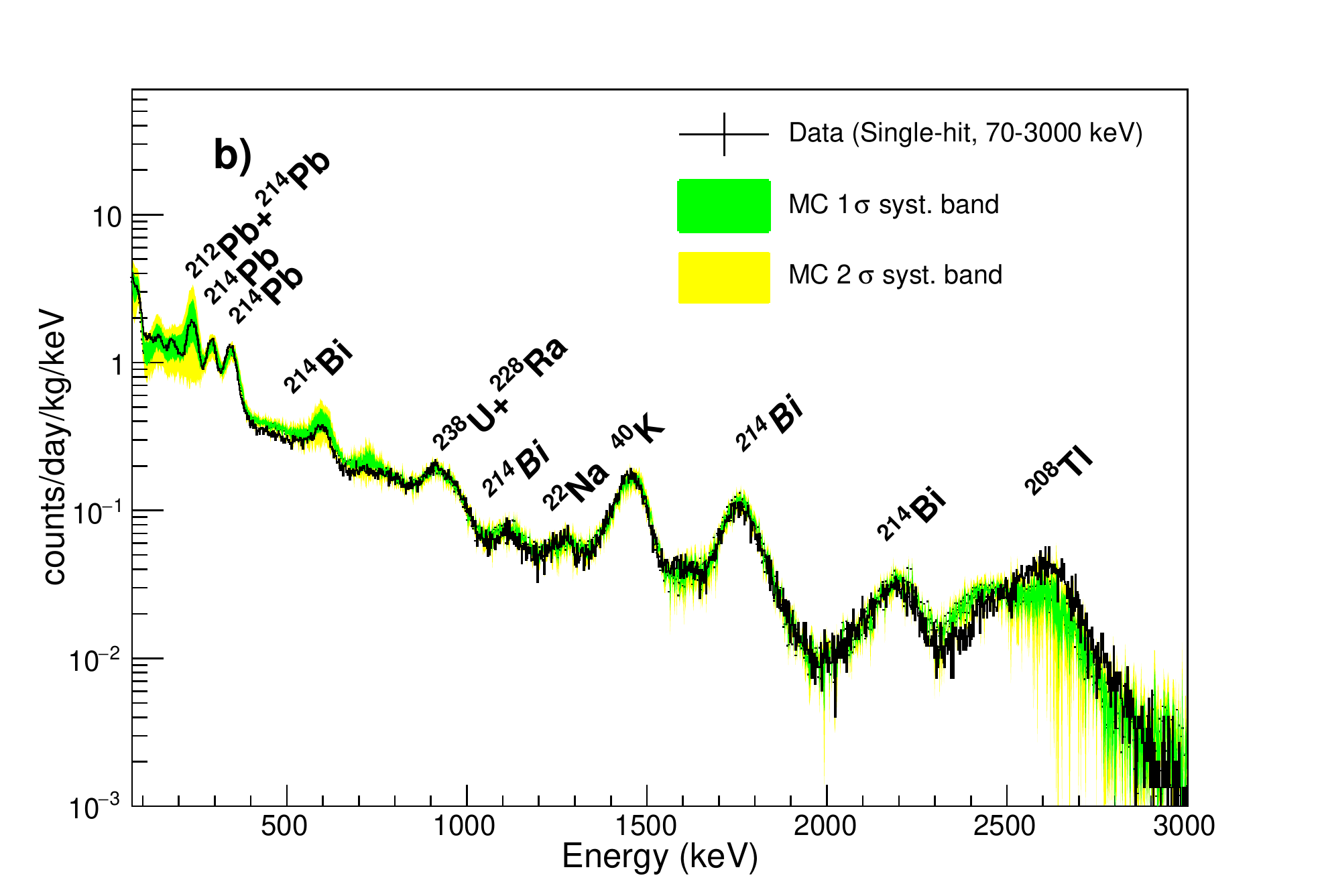}\\
\includegraphics[width=0.5\textwidth]{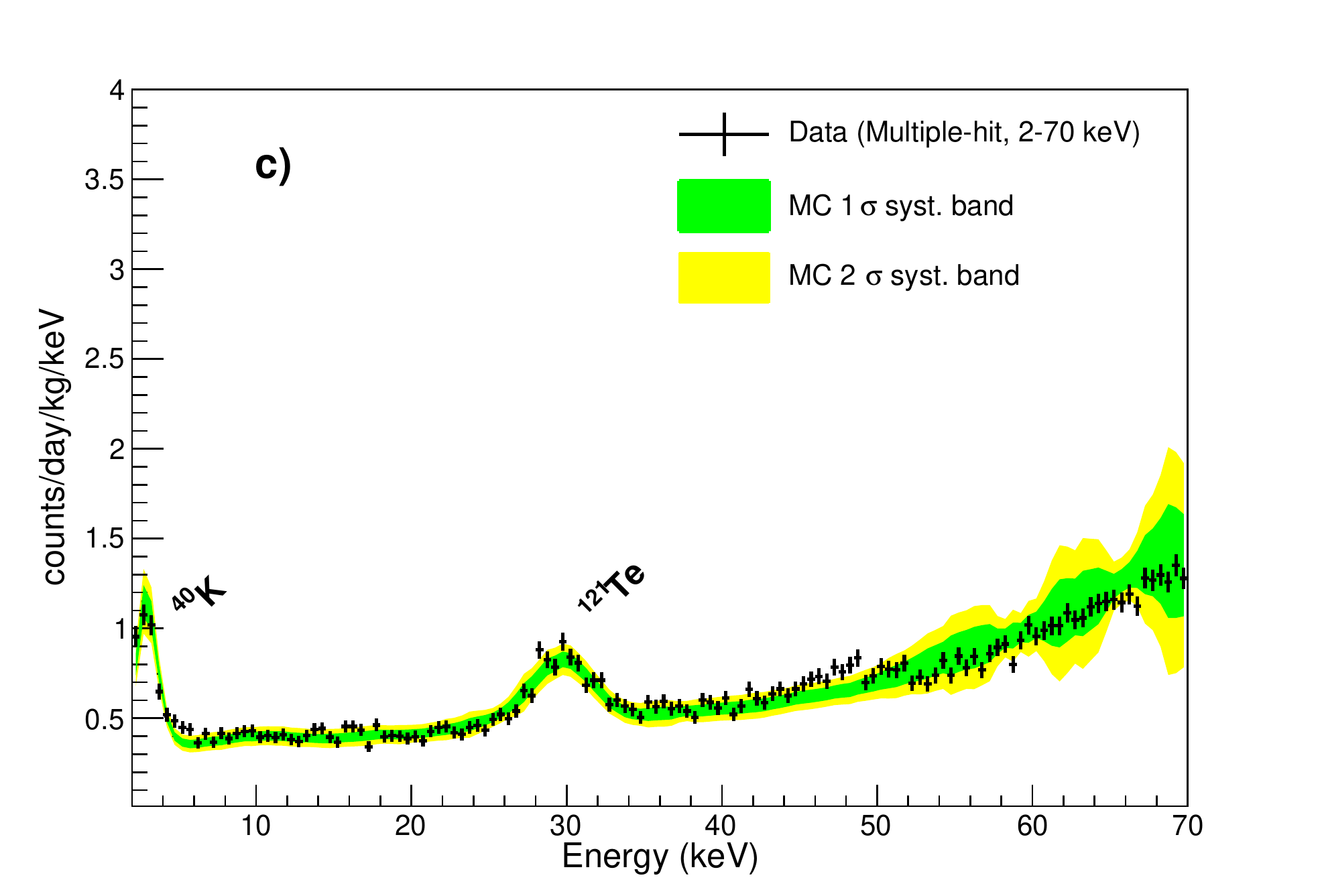}
\includegraphics[width=0.5\textwidth]{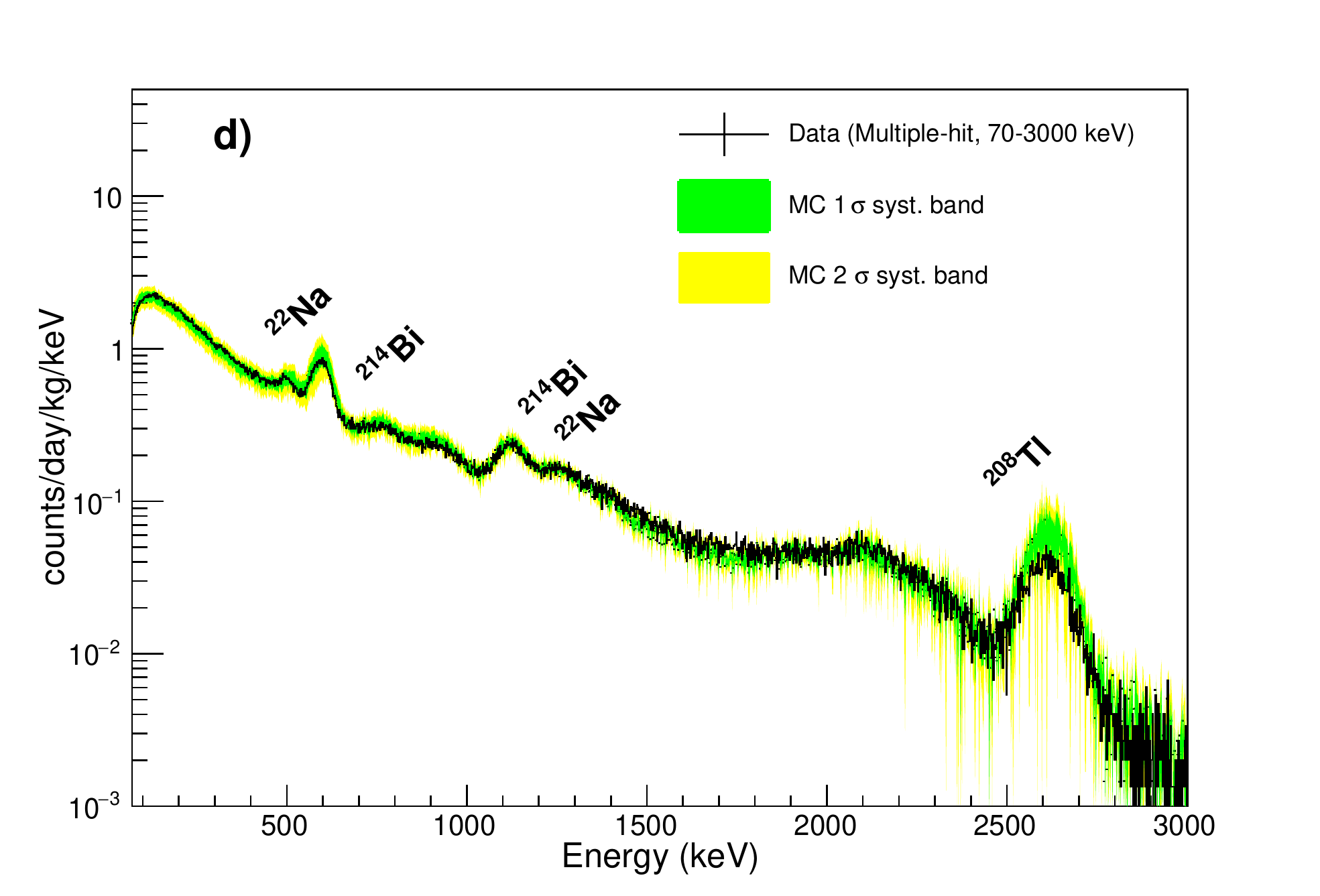}
\end{tabular}

\caption{
        {\bf A comparison between data and simulation.}
Four categories of data are shown: a) single-hit low-energy between 2--70\,keV;
b) single-hit between 70--3,000\,keV; c) multiple-hit between 2--70\,keV; d) multiple-hit between 70--3,000\,keV. The black points with 68\% CL error bars are data and
the green (yellow) band shows the $\pm1\sigma$ ($\pm2\sigma$) uncertainty range of the model.
The peak near 3\,keV in the multiple-hit, low-energy spectrum (panel c)) is due the tagged $^{40}$K events.
A small inset a--1) in panel a) shows a zoomed-in view in the region of interest after efficiency corrections are applied.
The major contributors to the radioactive background are labelled. 
}
\label{fig:coverage}
\end{figure*}

\begin{figure*}
\begin{tabular}{ccc}
\includegraphics[width=0.33\textwidth]{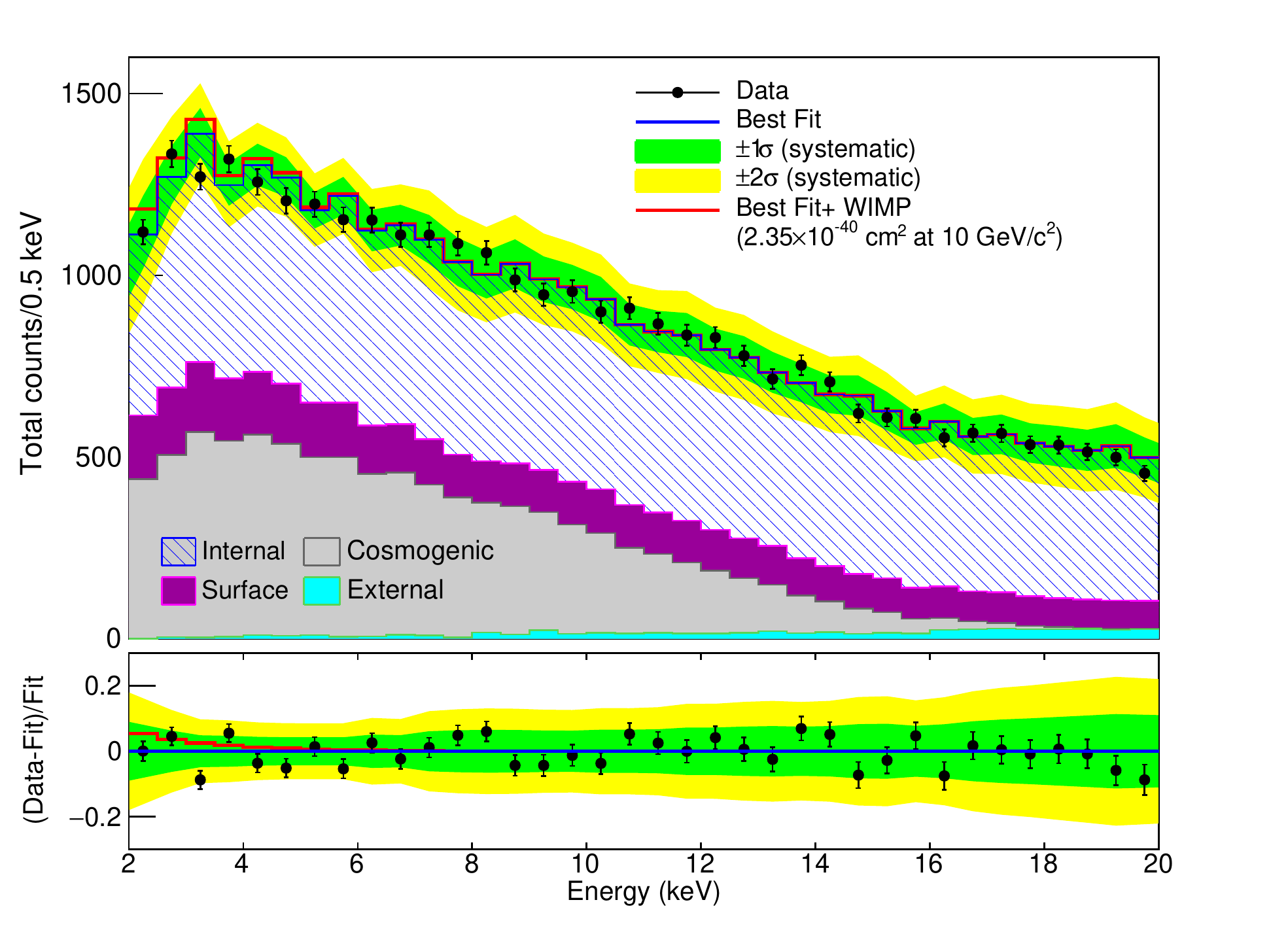} &
\includegraphics[width=0.33\textwidth]{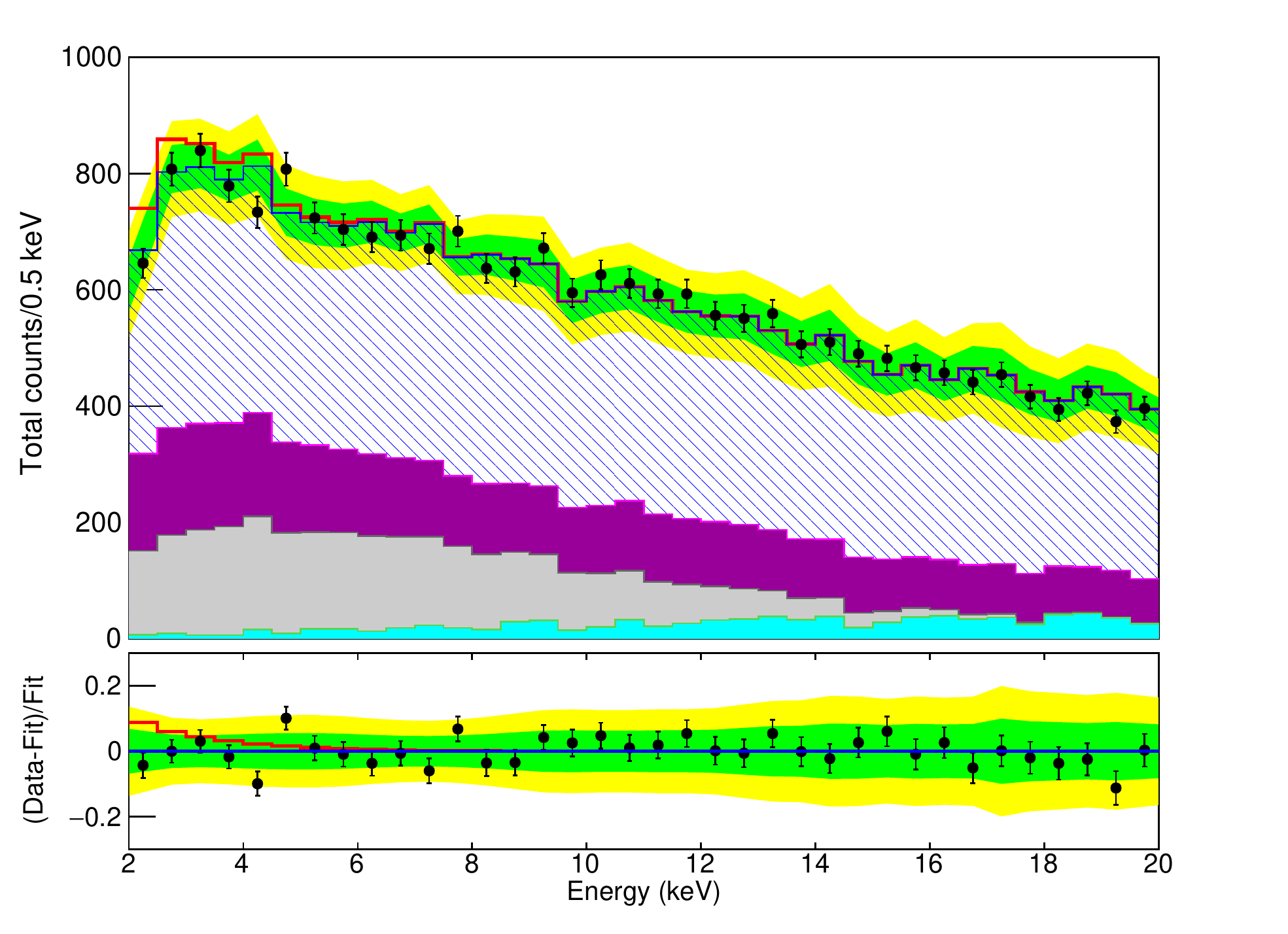} &
\includegraphics[width=0.33\textwidth]{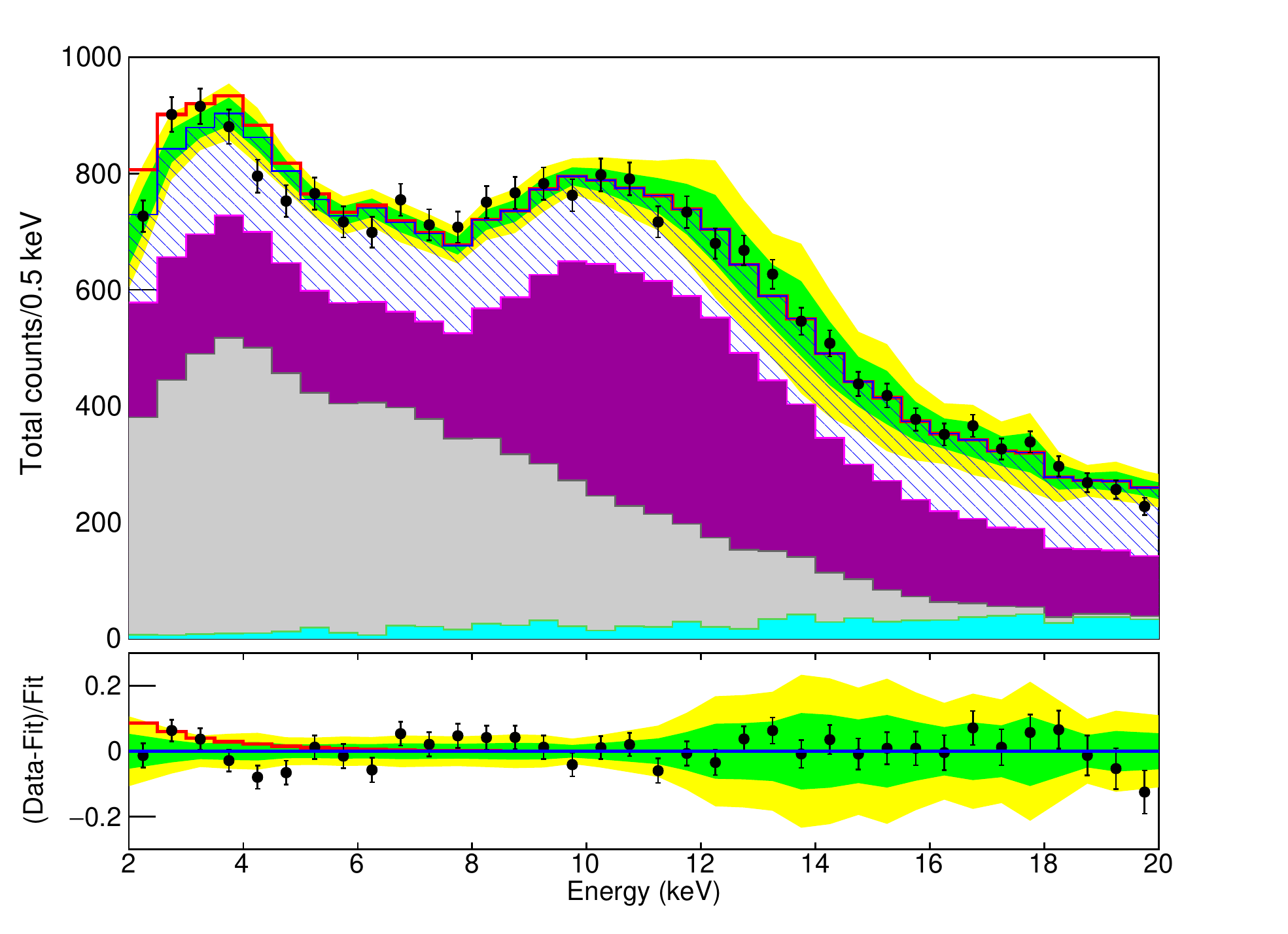} \\
a) Crystal 1 & b) Crystal 2 & c) Crystal 3 \\
\includegraphics[width=0.33\textwidth]{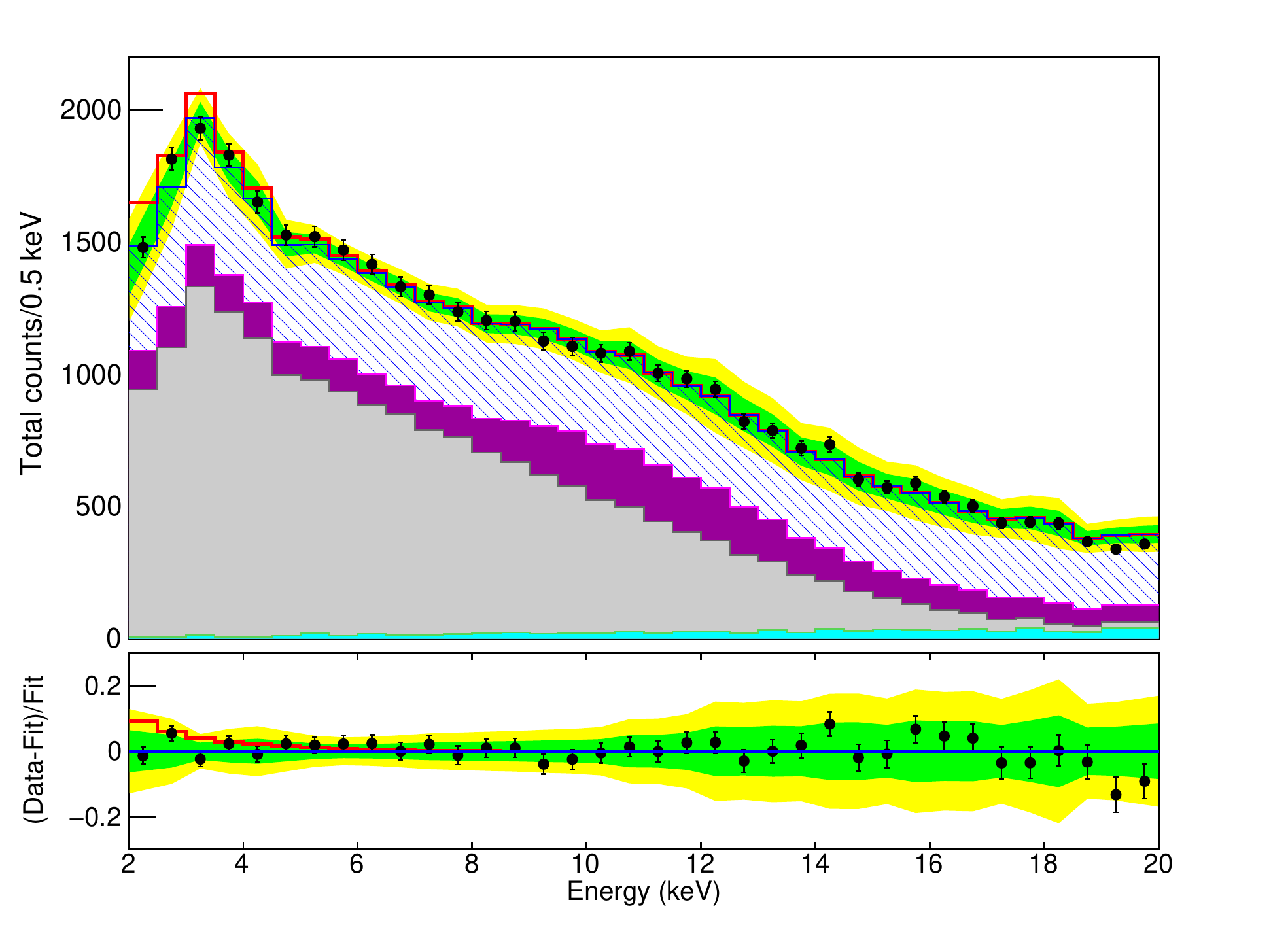} &
\includegraphics[width=0.33\textwidth]{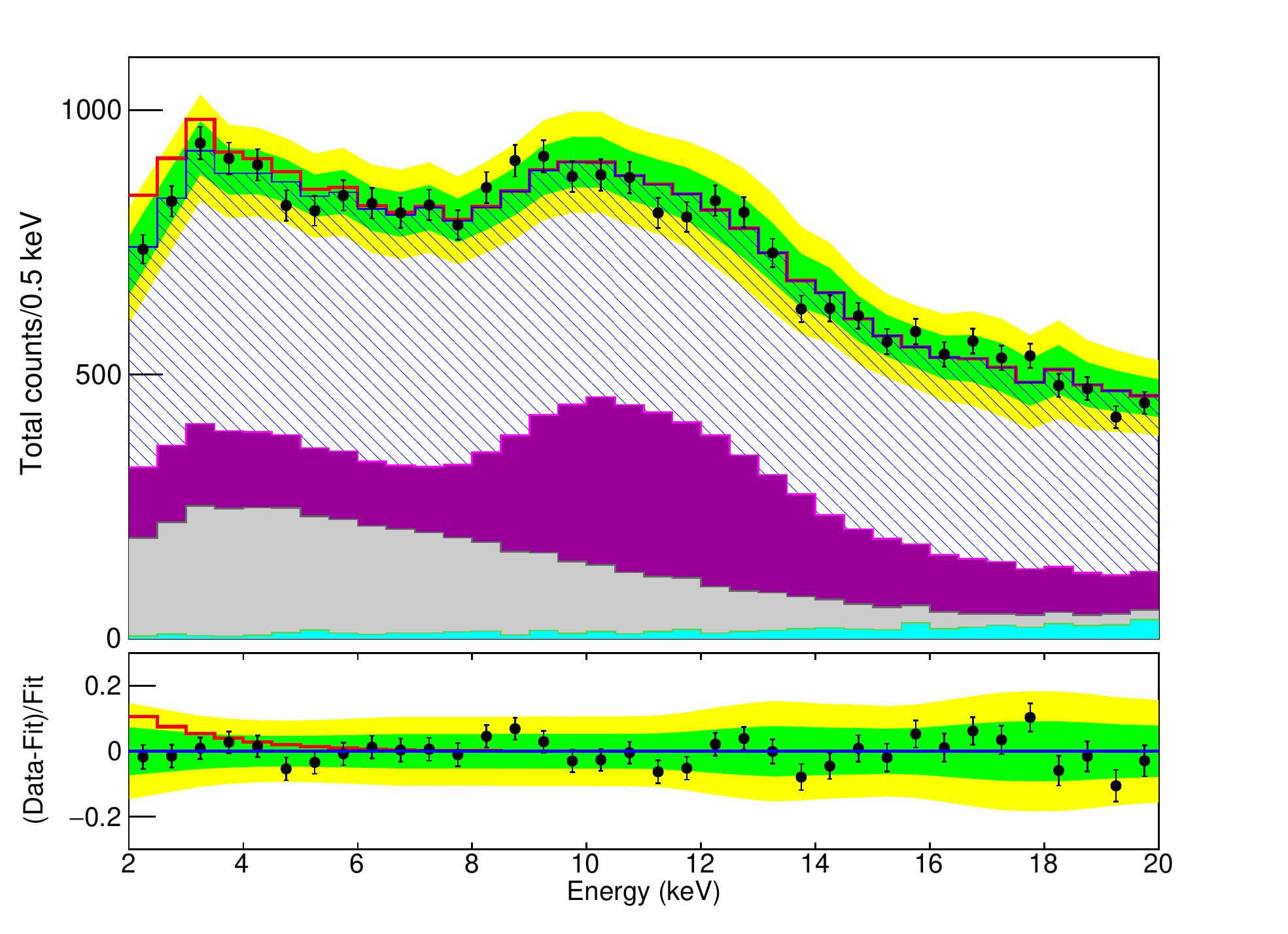} &
\includegraphics[width=0.33\textwidth]{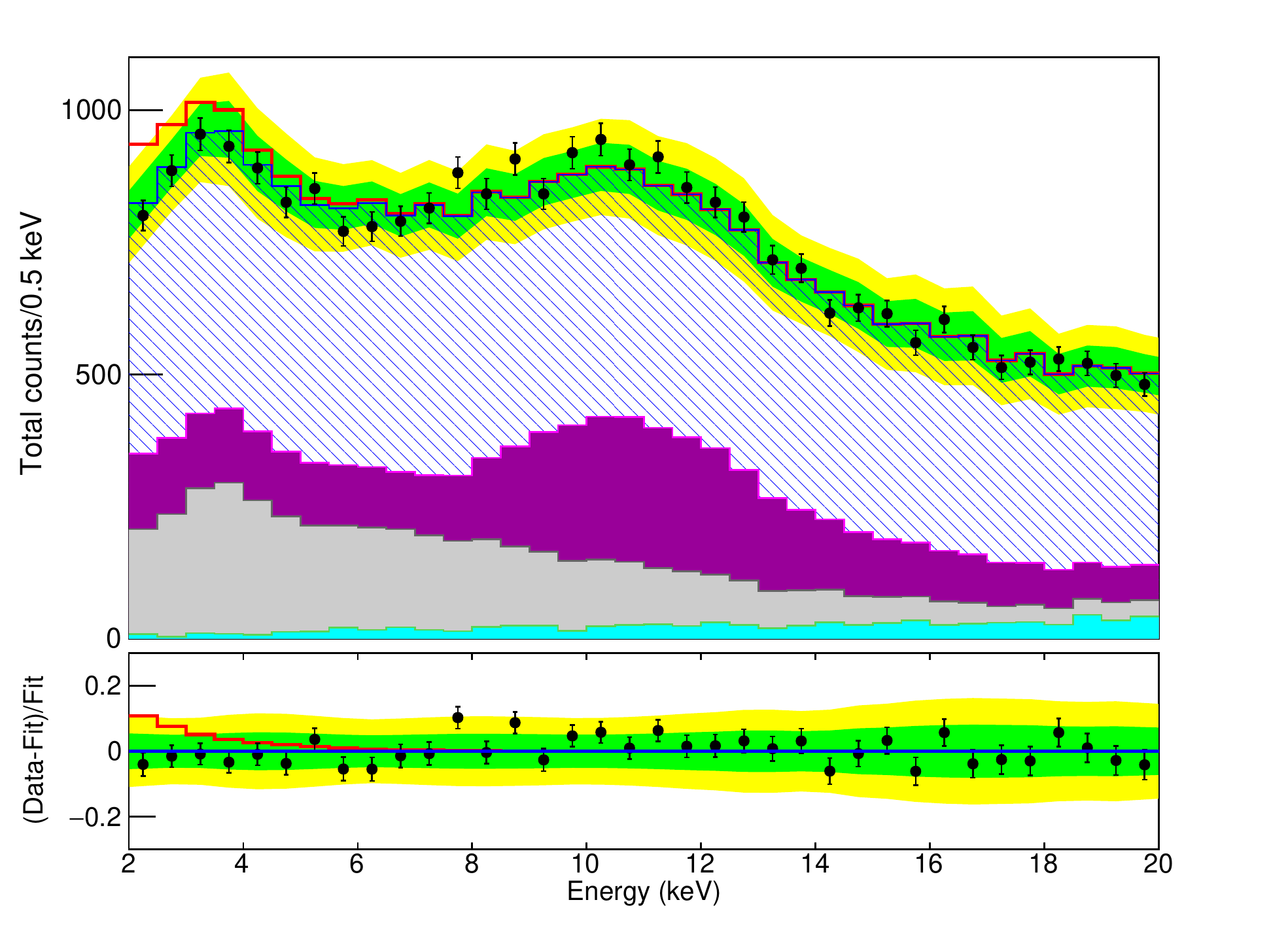} \\
d) Crystal 4 & e) Crystal 6 & f) Crystal 7  \\
\end{tabular}
\caption{
{\bf Crystal-by-crystal fit results.}
The points with 68\% CL error bars show the measured energy spectra for each of the six crystals a)--f). The fit results are shown as blue histograms with the $\pm 1\sigma$ ($\pm 2\sigma$) error bands shown in green (yellow).
To compare the signal strength of the DAMA-Na region with our data, a 10 GeV/c$^2$ WIMP signal at 2.35$\times$10$^{-40}$ cm$^2$ (DAMA-Na central region) is indicated for each crystal as a red histogram.
The fit residuals, together with the expectations for the 10~GeV/$c^2$ WIMP signal are also shown.
}
\label{fig:bestfitC}
\end{figure*}

\clearpage

\end{document}